\documentclass[%
 reprint,
showpacs,
 amsmath,amssymb,
aps,
prb,
floatfix,
]{revtex4-1}

\usepackage{graphicx}
\usepackage{dcolumn}
\usepackage{bm}
\usepackage[colorlinks = true, linkcolor = blue]{hyperref}
\usepackage[usenames,dvipsnames,svgnames,table]{xcolor}
\usepackage{placeins}

\usepackage[
text={7.5in,10in},centering,
]{geometry}

\usepackage{tabularx}
\usepackage{dcolumn}
\begin{document}


\title{Theory of proximity-induced exchange coupling in graphene on hBN/(Co, Ni)}

\author{Klaus Zollner, Martin Gmitra, Tobias Frank, and Jaroslav Fabian}
\affiliation{Institute for Theoretical Physics, University of Regensburg,
93040 Regensburg, Germany}

\date{\today}

\begin{abstract}
Graphene, being essentially a surface, can borrow some properties of an insulating substrate (such as exchange or spin-orbit couplings) while still preserving a great degree of autonomy of its electronic structure. Such derived properties are commonly labeled as proximity. 
Here we perform systematic first-principles calculations of the proximity exchange coupling, induced by cobalt (Co) and nickel (Ni) in graphene, via a few (up to three) layers of hexagonal boron nitride (hBN). We find that the induced spin splitting of the graphene bands is of the order of 10 meV for a monolayer of hBN, decreasing in magnitude but alternating in sign by adding each new insulating layer. We find that the proximity exchange can be giant if there is a resonant $d$ level of the transition metal close to the Dirac point.
Our calculations suggest that this effect could be present in Co heterostructures, in which a $d$ level strongly hybridizes
with the valence-band orbitals of graphene. Since this hybridization is spin dependent, the proximity spin splitting is unusually large, about 10 meV even for two layers of hBN. An external electric field can change the offset of the 
graphene and transition-metal orbitals and can lead to a reversal of the sign of the exchange parameter. This we predict to 
happen for the case of two monolayers of hBN, enabling electrical control of proximity spin polarization (but also 
spin injection) in graphene/hBN/Co structures. Nickel-based heterostructures show weaker proximity effects than cobalt heterostructures. We introduce two  phenomenological models to describe the 
first-principles data. The minimal model comprises the graphene (effective) $p_z$ orbitals and can be used to study transport
in graphene with proximity exchange, while the $p_z$-$d$ model also includes hybridization with $d$ orbitals, which is important
to capture the giant proximity exchange. Crucial to both models is the pseudospin-dependent exchange coupling, needed to 
describe the different spin splittings of the valence and conduction bands. 
\end{abstract}


\pacs{71.15.Mb, 73.22.Pr, 73.63.-b, 75.30.Et}
\maketitle

\section{Introduction}
\label{sec:Intro}
Graphene is a diamagnet with a weak
spin-orbit coupling, so spin interactions in devices containing clean graphene 
are rather weak \cite{Han2014}. One way to enhance these interactions is by functionalizing
graphene with adatoms and admolecules, which works well for both exchange \cite{yazyev2010, Yazyev2007, Wehling2007, McCreary2012, Mallet2016a, Ding2011} 
and spin-orbit 
\cite{PhysRevLett.103.026804,PhysRevLett.110.246602,Zhou20101405,
Irmer2014,PhysRevB.91.195408, PhysRevLett.110.156602, PhysRevX.1.021001, Balakrishnan2013, Zollner2015, Ding2011} couplings. 
Functionalized graphene has local ``hot spots" of giant 
exchange and spin-orbit fields, which can be used to investigate spin transport
\cite{Kochan2014, Ferreira2014, Wilhelm2015, Thomson2015, Roche2016, Bundesmann2015}.

A promising way to induce exchange coupling in graphene is placing it on a ferromagnetic 
substrate. In order to preserve the Dirac band structure, the substrate should be a ferromagnetic 
insulator, as in the density functional theory (DFT) study of graphene on EuO \cite{Yang2013}, predicting $20\%$ spin polarization
of graphene bands, an antiferromagnetic insulator \cite{Qiao2014}, or a ferromagnetic metal separated from graphene by an insulating barrier. The advantage of this
approach over functionalizing graphene with adatoms is that the induced band structure effects are uniform;
one can speak of a proximity electronic band structure with the hope of further electrical control. 

In fact, heterostructures of graphene with ferromagnets are essential for introducing 
spintronic phenomena \cite{RevModPhys.76.323,fabian2007semiconductor}
in graphene. Proximity exchange in graphene on a ferromagnetic insulator has recently 
been experimentally investigated for spin transport \cite{Wang2015, Leutenantsmeyer2016, Wei2016}, while tunnel junctions of graphene with ferromagnetic metals have been widely 
used in experimental demonstrations of electrical spin injection into graphene 
\cite{Tombros2007, Ohishi2007, Yamaguchi2014, Yamaguchi2012, Volmer2013, Volmer2015, Han2010, Patra2012, Kamalakar2014, Kamalakar2016, Wu2014}. The benefits
turn out to be mutual: graphene can protect ferromagnets from oxidation and yield large spin tunneling
signals in ferromagnet/graphene interfaces \cite{Martin2015, Dlubak2012}, 
in agreement with theory \cite{Karpan2007, Lazic2014}. It is also predicted that graphene on Co 
can strongly enhance the perpendicular magnetic anisotropy \cite{Yang2016}.

In this paper we present systematic first-principles investigations of the proximity exchange in graphene 
on a substrate comprising either Co or Ni and an hBN tunnel barrier. The barrier shields the Dirac bands from
strong hybridization with the metallic orbitals, but it also induces an orbital gap due to the
sublattice symmetry breaking \cite{Giovannetti2007}. 
The predicted band gap of a graphene/hBN structure is of the order of $50$~meV and may be essential for
building graphene-based field-effect transistor devices \cite{Britnell2012a}. The proximity of graphene and metals can 
lead to doping due to the different work functions and resulting charge transfer, as studied from first
principles in Refs. \onlinecite{Giovannetti2008, Frank2016}. A recent study has predicted that graphene on (hBN)/Co
can be effectively gated, and the induced spin polarization can be tuned by a transverse electric field \cite{Lazic2016}. 

We have studied tunnel structures of graphene and (Co, Ni) with up to three layers of hBN. For a single tunneling
layer, the proximity-induced exchange in the Dirac bands is about $10$~meV. The resulting spin splitting depends 
on the band (valence and conduction), which has motivated us to introduce a pseudospin-dependent exchange-coupling model of graphene's $p_z$ orbitals. This minimal model nicely explains the DFT data. As we increase
the number of hBN layers, the proximity exchange is expected to decrease exponentially. However, we observe
that in the case of two hBN layers on Co, the valence band of graphene remains spin split by about $10$~meV. 
This giant splitting is due to a Co $d$ orbital of energy close to the Dirac point that strongly couples to the
$p_z$ graphene orbitals. As this $d$ orbital is spin polarized, the resulting anticrossing of the corresponding
bands is seen as a giant spin splitting of $p_z$ orbitals. We then propose an extended effective model 
based on $p_z$ and $d$ orbitals to explain the hybridization-induced spin splitting. 
Certainly, this effect can be to a certain degree, an artifact of DFT, as
the $d$ levels need not be well described, so in the discussion we also look at the Hubbard $U$ effects
on the proximity structure. We indeed find that the strong hybridization is reduced. For $U=1$ the valence-band splitting reduces to about $5$~meV, which is still giant when compared with
the spin splitting of $0.2$~meV of the conduction band.
We therefore believe that this giant enhancement of proximity exchange in Co-based devices with two hBN 
layers could be observed. In the Ni-based structures that we studied this effect is absent.

As the number of hBN layers increases one by one, the proximity exchange changes sign. This
is reminiscent of the interlayer coupling in ferromagnet/metal/ferromagnet structures, in which the 
coupling strength between the two ferromagnets is an oscillating function of the spacer thickness 
\cite{Bruno1993}. Another means to change the proximity exchange is to apply a transverse electric field. 
However, we find that in most (of our investigated) cases mainly the orbital parameters (staggered potential
and Dirac point offset from the Fermi level) are affected. 
The exchange parameters change much less. 
The notable exception is the aforementioned slab of graphene and Co, with two hBN layers. Here the 
proximity exchange is strongly affected by the position of the $d$ orbitals of Co, so the electric field
leads to a strong modification of the induced spin polarization in graphene. We even observe a 
crossover at electric fields close to $2$~V/nm, where the sign of the exchange changes, making electrical control of the orientation of the (equilibrium) proximity spin polarization in graphene possible.
Finally, we investigate the proximity effects with respect to the number of ferromagnetic layers, finding
that three layers are already representative of the bulk. We also give magnitudes of the induced spin splitting
of the hBN valence and conduction bands, which are active in spin-dependent tunneling. 

Our investigations should be useful for interpreting spin injection and spin tunneling data in graphene/hBN/(Co, Ni)
devices. Especially in cases of thin tunnel barriers (one or two monolayers of hBN), there could be a sizable equilibrium
spin polarization in graphene underneath the ferromagnetic electrodes.  The proposed models could be used for simulations of spin transport in graphene with proximity
exchange. Overall, we find that Co is more interesting than Ni, as far as the proximity effects go, with Ni showing 
weaker proximity exchange effects and no signatures of giant spin-dependent hybridization between Ni $d$ orbitals
and graphene $p_z$ ones. 
 
This paper is organized as follows. In Sec. \ref{sec:Computational} we describe the computational methods
and the investigated graphene/hBN/Co structures in detail.
In Sec. \ref{sec:Model} we introduce two model Hamiltonians describing the proximity Dirac bands. In Sec. \ref{sec:Exchange} we present the DFT results for graphene/hBN/(Co, Ni) slabs. There we also discuss
the behavior of the proximity band structure in the presence of more hBN layers and in the presence of 
a transverse electric field, as well as effects of the Hubbard $U$ and the thickness of the ferromagnetic layer.

\section{Computational Methods and System Definition}
\label{sec:Computational}
To study proximity-induced exchange interaction in graphene we consider a graphene/insulator/ferromagnet heterostructure in a slab geometry. 
Electronic states were calculated using DFT \cite{PhysRev.136.B864} within the  \textsc{quantum espresso} suite \cite{Giannozzi2009}. Self-consistent calculations were performed with a $k$-point sampling of $120\times 120\times 1$, if not indicated otherwise, in order to get the correct Fermi energy of the metal and to  obtain an accurate description of bands in an energy window of $\pm 1$~eV around the Fermi level \footnote{An analysis on the importance of an accurate $k$-point sampling is included in Ref. \onlinecite{Bokdam2013}.}. 
Open-shell calculations provide the spin-polarized ground state.
We used an energy cutoff for charge density of $450$~Ry, and the kinetic-energy cutoff for wave functions was $100$~Ry for the scalar relativistic pseudopotential with the projector augmented-wave method \cite{PhysRevB.59.1758} with the Perdew-Burke-Ernzerhof exchange-correlation functional \cite{PhysRevLett.77.3865}.
For the relaxation of the heterostructures, we added van der Waals corrections \cite{Grimme2006} and used the Broyden-Fletcher-Goldfarb-Shanno quasi-Newton algorithm \cite{quasinewton}. 
In order to simulate quasi-two-dimensional systems a vacuum of $15$~${\textrm{\AA}}$ was used
together with dipole corrections \cite{Bengtsson1999} to avoid interactions between periodic images in our slab geometry.
To determine the interlayer distances, the atoms were allowed to relax in their $z$ positions (transverse to the layers) until all components of
all forces were reduced below $10^{-4}$~(Ry/$a_0$), where $a_0$ is the Bohr radius. 

Initial atomic structures were set up with the atomic simulation environment (ASE) \citep{ASE}, as follows.
The lattice constant of graphene is $a=2.46$~${\textrm{\AA}}$ \cite{CastroNeto2009a}, the one for hBN is $ a=2.504$~${\textrm{\AA}}$ \cite{PhysRevB.36.6105}, and the one for hcp cobalt is $a=2.507$~${\textrm{\AA}}$ \cite{PhysRevB.16.5068}. We fix an effective average lattice constant of $a=2.489$~${\textrm{\AA}}$ for this well-lattice-matched system, as a compromise to make the lattices commensurable and to keep the unit cell as small as possible.
The lattice of graphene is strained by only 1\%.

\begin{figure}[htb] 
 \centering 
 \includegraphics[width=0.49\textwidth]{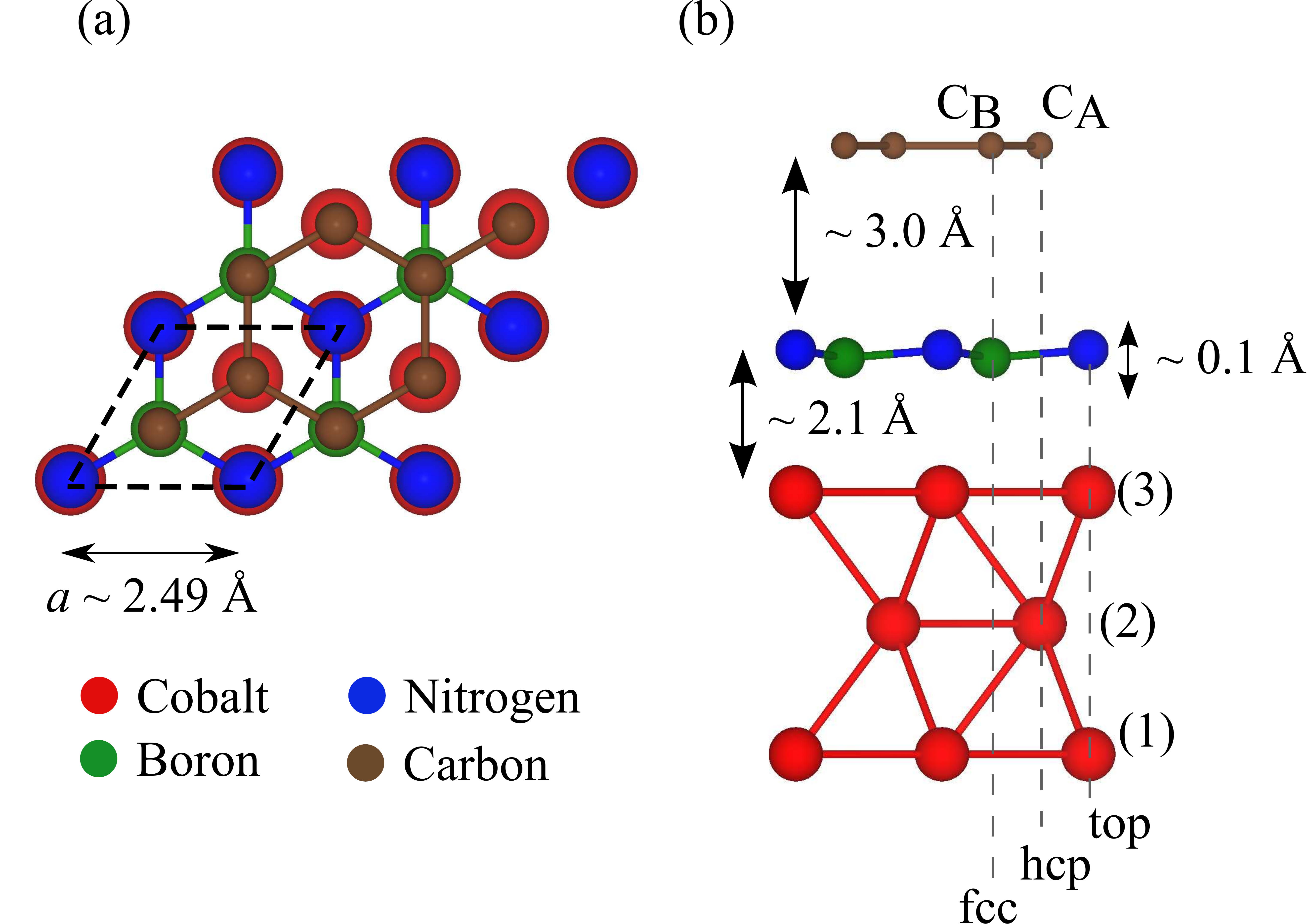}
 \caption{\label{fig:Structure_Co}(Color online) Structure of the graphene/hBN/Co system, with labels for the different atoms.
 (a) Top view of the structure, with one unit cell emphasized by the dashed line. (b) Side view with stacking configuration: $\textrm{C}_{\textrm{B}}$ is over boron, and $\textrm{C}_{\textrm{A}}$ is over hBN hexagon. Nitrogen is at the top site above Co, and boron is above the fcc site of Co. The distances indicated are measured between graphene/Co and the nitrogen atom of hBN, since the hBN layer is slightly corrugated by $\Delta z = 0.113$~${\textrm{\AA}}$. The boron atom is closer to the Co surface. Numbers in parenthesis indicate the Co layer.}  
 \end{figure}

We tested different stacking possibilities and found the energetically preferential structure. 
In Fig.~\ref{fig:Structure_Co}, we show our definition of the unit cell of the graphene/hBN/Co structure. A computational unit cell contains two carbon atoms, $\textrm{C}_{\textrm{A}}$ and $\textrm{C}_{\textrm{B}}$, forming graphene; one boron atom and one nitrogen atom per hBN layer; and three cobalt atoms, one per atomic layer. In general, three positions (top, hcp and fcc) can be distinguished within a hexagonal unit cell. 
The different positioning possibilities of the carbon atoms above the substrate  will influence the strength of the proximity magnetism. 
We get the lowest-energy configuration when nitrogen atoms are at the top sites and boron atoms are at the fcc sites above Co. 
Carbon atoms sit on top of boron atoms and at the hollow position above hBN (see Fig.~\ref{fig:Structure_Co}). 
These findings are in agreement with previous DFT studies \cite{Bokdam2013, Bokdam2014, Giovannetti2007}. 
After relaxation of atomic positions we obtained layer distances of $d_{\textrm{Co/hBN}} = 2.099$~${\textrm{\AA}}$~between the cobalt and hBN and $d_{\textrm{hBN/Gr}} = 3.010$~${\textrm{\AA}}$~between hBN and graphene (measured between C/Co and N atoms, respectively, since the hBN layer is corrugated).
The layer distances of this minimum energy configuration are roughly in agreement with those in Refs. \onlinecite{Giovannetti2007, Vanin2010, Bokdam2014}, which report $d_{\textrm{hBN/Gr}} = 3.22 - 3.40$~${\textrm{\AA}}$~and $d_{\textrm{Co/hBN}} = 1.92 - 2.02$~${\textrm{\AA}}$. 
We also find that the hBN layer is not flat anymore but slightly buckled since the boron atom is closer to the Co surface by $0.113$~${\textrm{\AA}}$~compared to the nitrogen atom, in agreement with Refs. \onlinecite{Bokdam2014, Grad2003}. 

For the stacking of hBN itself, when we use more than one monolayer of hBN, we use an AA$'$ stacking (B over N, N over B), which is the energetically favorable one, as shown in Ref. \onlinecite{Pease1952}, with distances between the layers in the range of $d_{\textrm{hBN/hBN}} = 2.98 - 3.09$~${\textrm{\AA}}$~(details are given in sections \ref{subsubsec:2hBN_Co} and \ref{subsubsec:3hBN_Co}).

\section{Effective Hamiltonian}
\label{sec:Model}
Our main goal is to answer the question, how do hBN and the ferromagnetic substrate affect the graphene Dirac cone at K?

In Fig. \ref{fig:Effects}(a) we show the calculated spin-resolved band structure of the graphene/hBN/Co system
(see Fig. \ref{fig:Structure_Co}), along the high-symmetry path M\textendash K\textendash $\Gamma$ in the energy window from $-5$ to $3$~eV. We can see that the linear dispersion of graphene around the K point is preserved and that the Dirac point is roughly $-0.5$~eV below the system Fermi level. 
Other bands, especially those in the vicinity of the Dirac point energy $E_{\rm D}$, originate mainly from the $d$ states of the ferromagnet which are located around the Fermi energy. 
The bands at the K point originating from hBN are far away from the Dirac point, with the highest- (lowest-) lying valence (conduction) band located at $-4$~eV ($2$~eV) away from the system Fermi level, emphasized by thicker lines in Fig. \ref{fig:Effects}(a). \textit{Moreover, hBN becomes spin polarized: its bands spin split by about} $0.5$~eV.

\subsection{Minimal \texorpdfstring{$p_z$}{TEXT} model}
We introduce a minimal Hamiltonian to describe the proximity-induced exchange spin splitting in graphene, similar
to earlier derivations of effective Hamiltonians for the proximity spin-orbit coupling in graphene on transition-metal dichalcogenides \cite{Gmitra2015, Gmitra2016} and on the Cu(111) substrate \cite{Frank2016}.  
\begin{figure}[htb] 
 \centering 
 \includegraphics[width=0.48\textwidth]{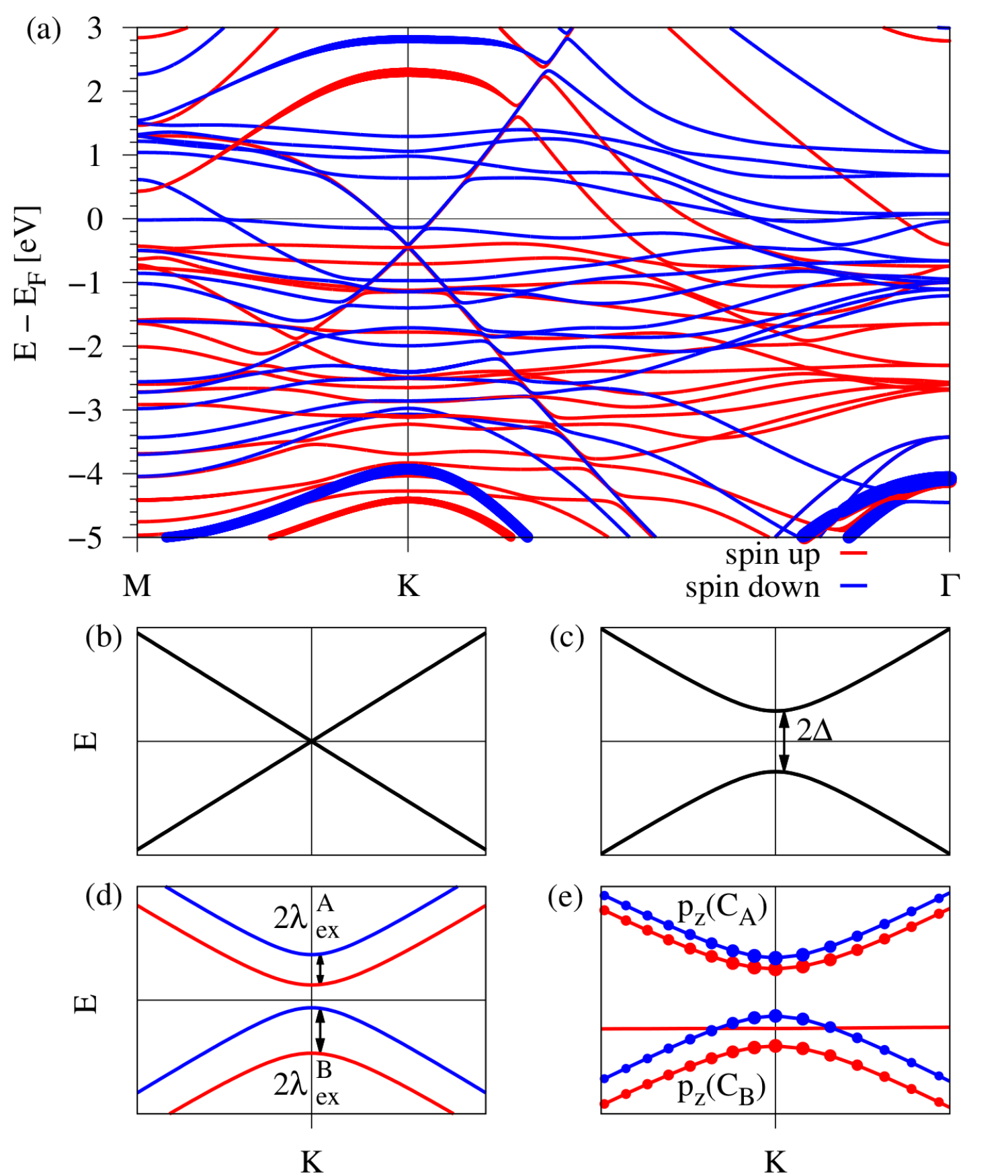}
 \caption{\label{fig:Effects}(Color online) Band structure of the graphene/hBN/Co system. (a) Calculated spin resolved band structure along the high-symmetry path M\textendash K\textendash $\Gamma$ using DFT. Spin-up (-down) states are shown in solid red (blue). Thicker bands at energies smaller (larger) than $-4$~eV ($2$~eV) correspond to the highest- (lowest-) lying valence (conduction) band of hBN which is spin split. (b) Hamiltonian $\mathcal{H}_{0}$ gives the linear dispersion of graphene with $v_{\textrm{F}}$ defining the slope. (c) $\mathcal{H}_{0}+\mathcal{H}_{\Delta}$ describes gapped graphene with a gap of $2\Delta$. (d) $\mathcal{H}_{0}+\mathcal{H}_{\Delta}+\mathcal{H}_{\textrm{ex}}$ describes gapped graphene in which the spin degeneracy of the conduction (valence) band gets lifted with a splitting of $2 \lambda_{\textrm{ex}}^{\textrm{A}}$ ($2 \lambda_{\textrm{ex}}^{\textrm{B}}$). (e) Close-up of  the band structure from (a) around the K point with labels for the different orbital and sublattice contributions of graphene; that is, the upper (lower) two bands are formed by $p_z$ orbitals of sublattice A (B).}  
\end{figure}

Pristine graphene is described by the massless Dirac Hamiltonian $\mathcal{H}_{0}$ in the vicinity of K (K'):
\begin{equation}
\mathcal{H}_{0} = \hbar v_{\textrm{F}} (\tau\sigma_{x}k_{x}+\sigma_{y}k_{y}),
\end{equation}
where $v_{\textrm{F}}$ denotes the Fermi velocity, $k_{x}$ and $k_{y}$ are the Cartesian components of the electron wave vector measured from K~(K'), and $\sigma_{x}$ and $\sigma_{y}$ are the pseudospin Pauli matrices acting on
the A and B sublattice orbitals. 
Hamiltonian $\mathcal{H}_{0}$ describes gapless Dirac states with conical dispersion near Dirac points, with $\tau = \pm 1$ for the K~(K') point, shown in Fig. \ref{fig:Effects}(b). 

Since graphene is on a hBN/Co substrate, the carbon atoms from different sublattices feel different potentials, leading to the Hamiltonian
\begin{equation}
\mathcal{H}_{\Delta} = \Delta \sigma_{z}s_{0},
\end{equation}
with $\sigma_{z}$ being the pseudospin Pauli matrix, $s_{0}$ being the unit spin matrix, and $\Delta$ being the proximity-induced orbital gap of the spectrum. The Hamiltonian $\mathcal{H}_{\Delta}$ describes a mass term, which breaks the pseudospin symmetry, and thus $\mathcal{H}_{0}+\mathcal{H}_{\Delta}$ describes a gapped graphene dispersion, shown in Fig. \ref{fig:Effects}(c).

To study the proximity exchange, we introduce the Hamiltonian
\begin{eqnarray}
\mathcal{H}_{\textrm{ex}} &&= \lambda_{\textrm{ex}}^{\textrm{A}} \left[(\sigma_{z}+\sigma_{0})/2\right]s_{z}\nonumber\\
&&+\lambda_{\textrm{ex}}^{\textrm{B}} \left[(\sigma_{z}-\sigma_{0})/2\right]s_{z}, 
\end{eqnarray}
with $\lambda_{\textrm{ex}}^{\textrm{A}}$ and $\lambda_{\textrm{ex}}^{\textrm{B}}$ being the exchange parameters for sublattices A and B, respectively. 
The dispersion of the Hamiltonian $\mathcal{H}_{0}+\mathcal{H}_{\Delta}+\mathcal{H}_{\textrm{ex}}$ is shown in Fig. \ref{fig:Effects}(d), along with the spin character.
This term is similar to a sublattice-resolved intrinsic spin-orbit coupling Hamiltonian \cite{Gmitra2015, Gmitra2016, Frank2016, PhysRevLett.110.246602, Irmer2014, Zollner2015}. The only difference is that it breaks time-reversal symmetry, associated with the magnetization.
A close-up in the vicinity of the K point of the DFT band structure is shown in Fig \ref{fig:Effects}(e), supporting
the need for a sublattice-resolved exchange Hamiltonian.
 
The proximity exchange (pex) Hamiltonian,
\begin{equation}
\label{eq:pex_Hamiltonian}
\mathcal{H}_{\rm pex} = \mathcal{H}_{0}+\mathcal{H}_{\Delta}+\mathcal{H}_{\textrm{ex}}
\end{equation}
is a minimal model using only effective carbon $p_z$ orbitals, which can be used to fit the DFT data directly at the K point and extract the pure band splittings. This Hamiltonian can be used for model charge and spin transport
calculations. The parameters $\Delta$, $\lambda_{\textrm{ex}}^{\textrm{A}}$, and $\lambda_{\textrm{ex}}^{\textrm{B}}$ are related to the band splittings at the K point: splitting of the conduction bands $\Delta E_{\rm cond} = |2\lambda_{\textrm{ex}}^{\textrm{A}}|$, splitting of the valence bands
$\Delta E_{\rm val} = |2\lambda_{\textrm{ex}}^{\textrm{B}}|$, and orbital gap
$\Delta E_{\rm cond-val} = |2\Delta|$,
as shown in Figs. \ref{fig:Effects}(c) and \ref{fig:Effects}(d).

The Dirac point is shifted in energy with respect to the system Fermi level by an energy $E_{\textrm{D}}$, as can be seen in Fig. \ref{fig:Effects}(a). We call the energy $E_{\textrm{D}}$ the Dirac point energy, and it will be our measure for the doping level.
The Dirac point energy $E_{\textrm{D}}$ for this model is calculated by averaging the four DFT energies of the graphene Dirac bands at the K point. 

\subsection{Extended \texorpdfstring{$p_z$-$d$}{TEXT} Hamiltonian}
The DFT results in Fig. \ref{fig:Effects}(a) show that in the interesting range of energies of the graphene Dirac bands, there can
also lie flat $d$ orbitals from Co. When these orbitals hybridize with $p_z$ carbon orbitals in graphene, the effective 
exchange coupling gets strongly modified. In order to capture this effect quantitatively, we extend our minimal 
model by a set of $d$ orbitals that appear close to the Dirac point. Similar effects occur in graphene on the Cu(111) substrate\cite{Frank2016}, for example. 

Figure \ref{fig:Effects}(a) shows that in the energy window of $\pm 400$~meV from the Dirac point, three $d$ bands interact with the Dirac states. We then extend our model by adding an effective ferromagnet Hamiltonian $\mathcal{H}_{\textrm{FM}}$, consisting of three $d$ bands, which describes the hybridization of the ferromagnet $d$ bands with the graphene states. The full effective Hamiltonian in the basis $|\textrm{A}\uparrow\rangle$, $|\textrm{A}\downarrow\rangle$, $|\textrm{B}\uparrow\rangle$, $|\textrm{B}\downarrow\rangle, \textrm{u}, \textrm{v}, \textrm{w}$, 
where $\textrm{u}$, $\textrm{v}$, and $\textrm{w}$ label the three $d$ orbitals (of the spin specified by the DFT), reads
\begin{widetext}
\begin{equation}
\label{eq:Exchange_Hamiltonian}
\mathcal{H}_{p_z\textrm{-}d} = \mathcal{H}_{0}+\mathcal{H}_{\Delta}+\mathcal{H}_{\textrm{ex}}+\mathcal{H}_{\textrm{FM}}
=
\begin{pmatrix}
\tilde{\Delta}+\tilde{\lambda}_{\textrm{ex}}^{\textrm{A}} & 0 & \hbar v_{\textrm{F}}(k_x-\textrm{i}k_y) & 0 & \textrm{u}_{\uparrow}^{\textrm{A}} & \textrm{v}_{\uparrow}^{\textrm{A}} & \textrm{w}_{\uparrow}^{\textrm{A}}\\
0 & \tilde{\Delta}-\tilde{\lambda}_{\textrm{ex}}^{\textrm{A}} & 0 & \hbar v_{\textrm{F}}(k_x-\textrm{i}k_y) & \textrm{u}_{\downarrow}^{\textrm{A}} & \textrm{v}_{\downarrow}^{\textrm{A}} & \textrm{w}_{\downarrow}^{\textrm{A}}\\
\hbar v_{\textrm{F}}(k_x+\textrm{i}k_y) & 0 & -\tilde{\Delta}-\tilde{\lambda}_{\textrm{ex}}^{\textrm{B}} & 0 & \textrm{u}_{\uparrow}^{\textrm{B}} & \textrm{v}_{\uparrow}^{\textrm{B}} & \textrm{w}_{\uparrow}^{\textrm{B}}\\
0 & \hbar v_{\textrm{F}}(k_x+\textrm{i}k_y) & 0 & -\tilde{\Delta}+\tilde{\lambda}_{\textrm{ex}}^{\textrm{B}} & \textrm{u}_{\downarrow}^{\textrm{B}} & \textrm{v}_{\downarrow}^{\textrm{B}} & \textrm{w}_{\downarrow}^{\textrm{B}}\\
\textrm{u}_{\uparrow}^{\textrm{A}} & \textrm{u}_{\downarrow}^{\textrm{A}} & \textrm{u}_{\uparrow}^{\textrm{B}} & \textrm{u}_{\downarrow}^{\textrm{B}} & E_{\textrm{u}} & 0 & 0\\
\textrm{v}_{\uparrow}^{\textrm{A}} & \textrm{v}_{\downarrow}^{\textrm{A}} & \textrm{v}_{\uparrow}^{\textrm{B}} & \textrm{v}_{\downarrow}^{\textrm{B}} & 0 & E_{\textrm{v}} & 0\\
\textrm{w}_{\uparrow}^{\textrm{A}} & \textrm{w}_{\downarrow}^{\textrm{A}} & \textrm{w}_{\uparrow}^{\textrm{B}} & \textrm{w}_{\downarrow}^{\textrm{B}} & 0 & 0 & E_{\textrm{w}} \\
\end{pmatrix}.
\end{equation}
\end{widetext}
Here $E_{j}$, $j = \textrm{u,v,w}$, are the energies that correspond to the $d$ states, and ${j}_{\uparrow, \downarrow}^{\textrm{A,B}}$ are the effective hybridization parameters with the corresponding Dirac state, where the subscript (superscript) indicates the spin (pseudospin) state. The proximity exchange ($\tilde{\lambda}_{\rm ex}$)
and orbital gap ($\tilde{\Delta}$) parameters are, in principle, different from those of the minimal Hamiltonian
$\mathcal{H}_{\rm pex}$, as they are renormalized due to the hybridization. 
The hybridization parameters vanish if the Dirac states directly at K are relatively far from the $d$ bands. 
As soon as the Dirac states at K are close to a $d$ band or the hybridization is so large that it affects the Dirac states, we can describe the hybridization with the corresponding interaction parameter. 
We call the Hamiltonian $\mathcal{H}_{p_z\textrm{-}d}$ the $p_z$-$d$-model. The energy window of roughly $\pm 150$~meV from the Dirac point energy can be described by this model rather well (see Figs. \ref{fig:Spin_bands_Co_1hBN_Gr} and \ref{fig:Spin_bands_Co_2hBN_Gr}). 

The $p_z$-$d$ model describes the hybridization between the surface $d$ bands of Co with the graphene Dirac states. 
As we will discuss, this hybridization significantly enhances the effective proximity exchange splitting.
Like the minimal model $\mathcal{H}_{\rm pex}$, the $p_z$-$d$ model has to be shifted in energy to match the DFT data. 
We call this energy $E_0$, and it is an analog of $E_\textrm{D}$. 

\section{Proximity exchange interaction}
\label{sec:Exchange}

We present our DFT calculations of the proximity exchange in graphene/hBN/Co and graphene/hBN/Ni for one, two, and 
three layers of hBN, as well as fits to the effective Hamiltonians. The proximity exchange decreases in magnitude but oscillates as the number of layers changes by one. This oscillating behavior is reminiscent 
of the oscillatory magnetic interlayer coupling \cite{Bruno1992, Bruno1993}.

\subsection{Graphene/hBN/cobalt}
\subsubsection{One hBN layer}
\begin{figure*}[htb] 
 \centering 
 \includegraphics[width=0.96\textwidth]{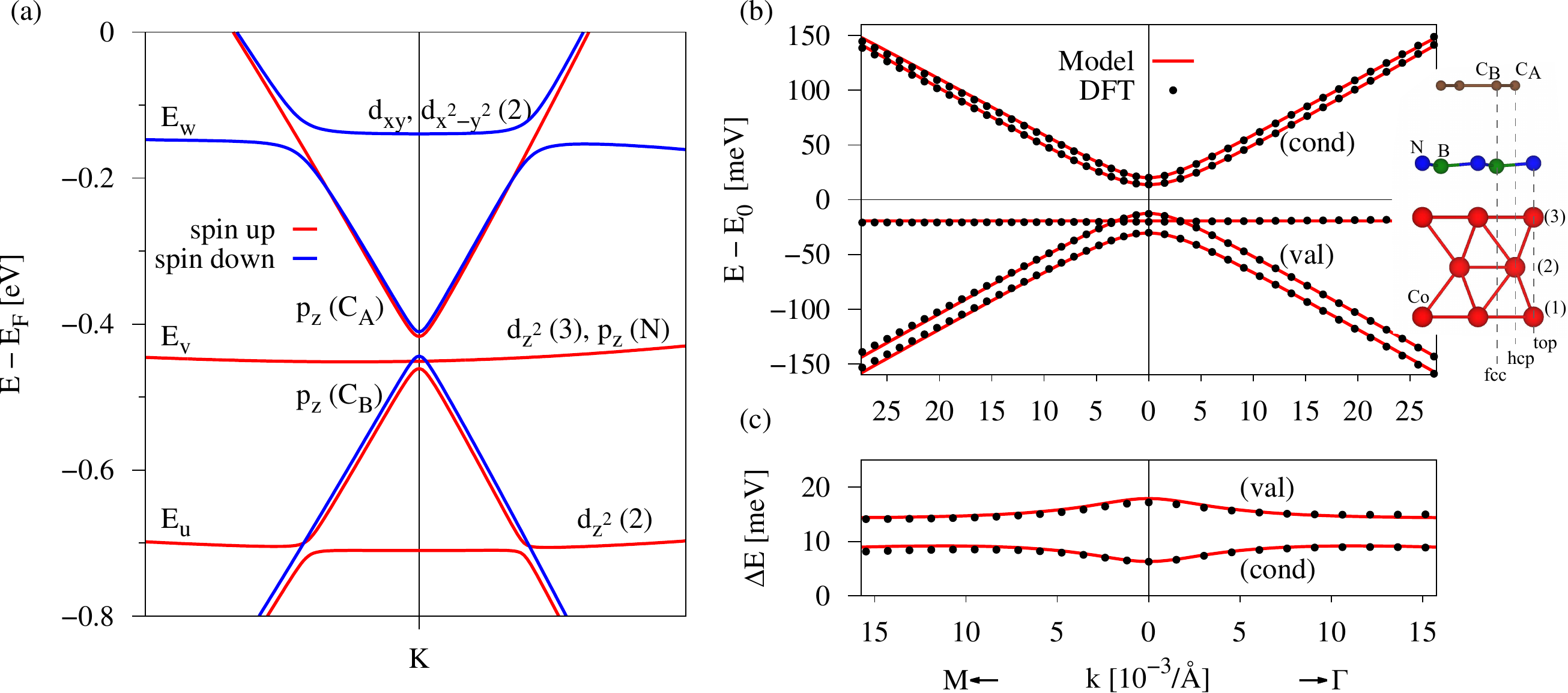}
 \caption{\label{fig:Spin_bands_Co_1hBN_Gr}(Color online) Calculated spin-polarized band structure of the graphene/hBN/Co heterostructure for one layer of hBN. (a) Band structure in the vicinity of the Dirac point with labels for the main orbital contributions from which the individual bands are formed; for example, $d_{z^2} (3)$ corresponds to the $d_{z^2}$ orbital of Co atom (3) from Fig. \ref{fig:Structure_Co}. Labels $E_{j}$, $j = \textrm{u,v,w}$, are the energy bands, which correspond to the Co $d$ states used to fit the $p_z$-$d$ model Hamiltonian in Eq. (\ref{eq:Exchange_Hamiltonian}). The energies $E_{j}$, in Eq. (\ref{eq:Exchange_Hamiltonian}), are measured with respect to the energy $E_{0}$. (b) The fit to the $p_z$-$d$ model with a side view of the structure. First-principles data (dotted lines) are well reproduced by the $p_z$-$d$ model (solid lines). (c) The corresponding splittings of the valence (val) and conduction (cond) Dirac states of graphene. The main fit parameters are $E_{0} = -430.89$~meV, $\tilde{\Delta} = 21.45$~meV, $\tilde{\lambda}_{\textrm{ex}}^{\textrm{A}} = -7.63$~meV, $\tilde{\lambda}_{\textrm{ex}}^{\textrm{B}} = 8.95$~meV, $E_{\textrm{u}} = -279.41$~meV, $E_{\textrm{v}} = -19.37$~meV, $E_{\textrm{w}} = 282.31$~meV, $\textrm{w}_{\downarrow}^{\textrm{A}}= 48.44$~meV.
The most relevant parameters are obtained by a least-squares fit, minimizing the difference between the model and the DFT data for a fitting range from K towards the $\Gamma$ point for $k$ points up to $20\times 10^{-3}/\textrm{\AA}$. By performing the fit towards the M point, one would obtain slightly different (by at most 5\%) parameters. 
From the band structure and from the fact that we limit our fitting range, we find that no additional hybridization parameters ${j}_{\uparrow, \downarrow}^{\textrm{A,B}}$ are necessary to fit our band structure, except for the mentioned ones. 
The Fermi velocity to match the slope away from the K point is $v_{\textrm{F}} = 0.812\times 10^6$~m/s, which corresponds to a nearest-neighbor hopping parameter of $t=2.48$~eV, slightly smaller than the commonly used value of $2.6$~eV \citep{Zollner2015,Irmer2014,PhysRevLett.110.246602} due to the larger lattice constant used here. }  
 \end{figure*}

Figure \ref{fig:Spin_bands_Co_1hBN_Gr}(a) shows the spin-polarized band structure of the graphene/hBN/Co heterostructure for one layer of hBN.
The graphene Dirac states for \textit{spin-up} are lying lower in energy than the \textit{spin-down} ones.
The Dirac point energy is below the system Fermi level, corresponding to electron doping of graphene, since the Fermi level now crosses the conduction band of graphene. 
This shift is induced by the metal, as already suggested in Ref. \onlinecite{Giovannetti2008}. 
The Co bands hybridize with the graphene states in the vicinity of the K point and introduce exchange splitting.

From the band structure in Fig.~\ref{fig:Spin_bands_Co_1hBN_Gr} we see that the linear dispersion of graphene is preserved.
In addition, a gap forms, and the spin degeneracy of the Dirac states gets lifted, allowing for semiconducting properties along with the usage of different spin channels by appropriate experimental setups. 
By comparing our DFT results to the $p_z$-$d$ model Hamiltonian $\mathcal{H}_{p_z\textrm{-}d}$, Eq. (\ref{eq:Exchange_Hamiltonian}), we obtain the parameters given in Table \ref{tab:summary} for Co as the ferromagnet and one layer of hBN.
The fit of the $p_z$-$d$ model is shown in Fig.~\ref{fig:Spin_bands_Co_1hBN_Gr}(b) and agrees very well with the DFT data. 
The gap in the dispersion is found to be roughly $40$~meV, while the band splittings are of the order of $10$~meV for one layer of hBN.

We additionally employ the minimal $\mathcal{H}_{\rm pex}$ model, Eq. (\ref{eq:pex_Hamiltonian}), valid directly at the K point. 
The parameters for the minimal $p_z$ model are given in Table \ref{tab:summary_pz}. The two models yield quantitatively very similar exchange parameters due to the rather weak hybridization of the $d$ orbitals with graphene ones. 

\subsubsection{Two hBN layers}
\label{subsubsec:2hBN_Co}
\begin{figure*}[htb] 
 \centering 
 \includegraphics[width=0.96\textwidth]{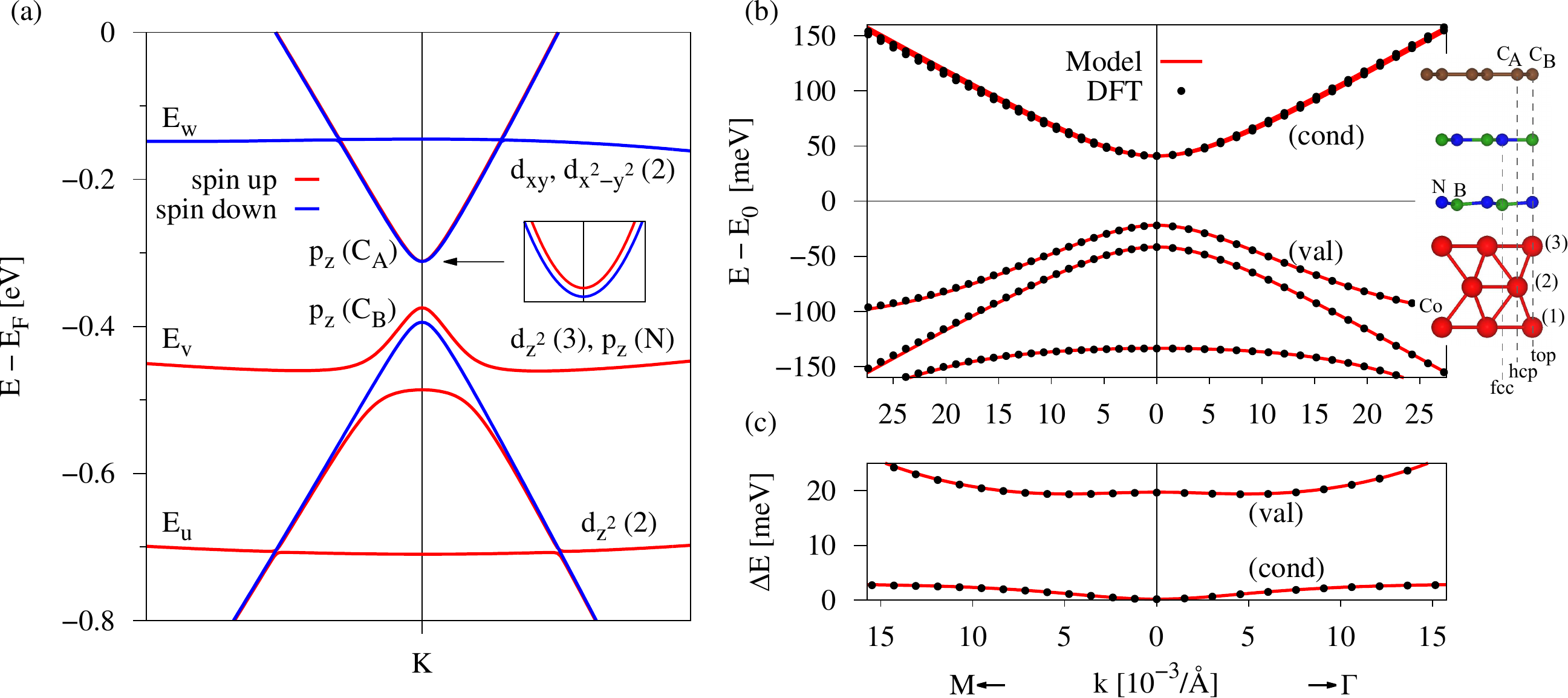}
 \caption{\label{fig:Spin_bands_Co_2hBN_Gr}(Color online) Spin-polarized band structure of the graphene/hBN/Co heterostructure for two layers of hBN (AA$'$ stacking). (a) Band structure in the vicinity of the Dirac point with labels for the main orbital contributions. The inset shows a close-up of the conduction Dirac states to visualize the reversal of the spin states. (b) The fit to the $p_z$-$d$ model with a side view of the structure for two layers of hBN. DFT data (dotted lines) are well reproduced by the $p_z$-$d$ model (solid lines). (c) The corresponding splittings of the valence and conduction Dirac states. The fit parameters are $ E_{0}= -352.65$~meV, $\tilde{\Delta} = 41.02$~meV, $\tilde{\lambda}_{\textrm{ex}}^{\textrm{A}} = 0.096$~meV, $\tilde{\lambda}_{\textrm{ex}}^{\textrm{B}} = -0.512$~meV, $E_{\textrm{u}} = -357.12$~meV, $E_{\textrm{v}} = -114.75$~meV, $E_{\textrm{w}} = 207.34$~meV, $\textrm{v}_{\uparrow}^{\textrm{B}}= 41.67$~meV.
The Fermi velocity to match the slope away from the K point is $v_{\textrm{F}} = 0.820\times 10^6$~m/s. All other parameters are zero for the same fitting range as for the one-layer case.}  
 \end{figure*}
Figure \ref{fig:Spin_bands_Co_2hBN_Gr} shows the calculated band structure and the fit to the $p_z$-$d$ model in the case of two layers of hBN and three layers of Co. The inset in Fig. \ref{fig:Spin_bands_Co_2hBN_Gr}(b) shows the geometry for two layers of hBN. 
The relative position of carbon atom $\textrm{C}_{\textrm{A}}$ to hBN is not changed, while the position of atom $\textrm{C}_{\textrm{B}}$ is changed, such that it is again on top of the uppermost boron atom, which is the energetically favorable situation for graphene on hBN. 
The conduction (valence) Dirac states are still formed by sublattice A (B), even though $\textrm{C}_{\textrm{B}}$ has changed its position within the unit cell. The layer distance between the two hBN layers was relaxed to
$d_{\textrm{hBN/hBN}} = 2.977$~${\textrm{\AA}}$, and the distance between the uppermost hBN layer and graphene is $d_{\textrm{hBN/Gr}} = 3.114$~${\textrm{\AA}}$ in the two-hBN-layer case. The corrugation of the lower hBN and the distance between hBN and Co did not change.
\begin{table*}[htb]
\begin{ruledtabular}
\begin{tabular}{lcccccccccc}

FM&
hBN&
\textrm{$E_{0}$ }&
\textrm{$\tilde{\Delta}$ }&
\textrm{$\tilde{\lambda}_{\textrm{ex}}^{\textrm{A}}$ }&
\textrm{$\tilde{\lambda}_{\textrm{ex}}^{\textrm{B}}$ }&
\textrm{$E_{\textrm{u}}$ }&
\textrm{$E_{\textrm{v}}$ }&
\textrm{$E_{\textrm{w}}$ }&
IA &
\textrm{$v_{\textrm{F}}/10^5$ } \\

 & (layers) & (meV)& (meV)& (meV)& (meV)& (meV)& (meV)& (meV)& (meV)& (m/s)\\
\colrule

Co & 1 & -430.89 & 21.45 & -7.63 & 8.95 &-279.41 &-19.37 & 282.31 & 48.44 ($\textrm{w}_{\downarrow}^{\textrm{A}}$) & 8.12\\
   & 2 & -352.65 & 41.02 & 0.096 & -0.512 & -357.12 & -114.75 & 207.34 & 41.67 ($\textrm{v}_{\uparrow}^{\textrm{B}}) $ & 8.20\\
   & 3 & -301.07 & 38.83 &-0.005 & 0.018 & -408.70 &-166.03 &155.10 &  & 8.21\\
   
   Ni & 1 & -527.98 & 22.98 & -1.25 & 8.17 & -272.58 & -201.27 & -158.17 & 10.15 ($\textrm{v}_{\uparrow}^{\textrm{B}})$, 4.19 ($\textrm{w}_{\downarrow}^{\textrm{B}}$) & 8.10\\
      & 2 &  -435.76 & 42.88 & 0.080 & -1.44 & -363.36 & -309.27 & -251.05 & 32.67 ($\textrm{v}_{\uparrow}^{\textrm{B}}) $& 8.24\\
      & 3 & -361.54 & 40.42 & -0.005 & 0.017 & -437.30 & -384.53 & -324.91 &  &8.26
\end{tabular}
\end{ruledtabular}
\caption{\label{tab:summary}
Summary of the most relevant parameters for all relevant structures ($a=2.489$~$\textrm{\AA}$ and $U = 0$~eV) for the different ferromagnets (FM) Co and Ni for one to three layers of hBN: proximity gap $\tilde{\Delta}$; energy shift $E_{0}$; exchange parameters $\tilde{\lambda}_{\textrm{ex}}^{\textrm{A}}$ and $\tilde{\lambda}_{\textrm{ex}}^{\textrm{B}}$; energies $E_{\textrm{u}}$, $E_{\textrm{v}}$, and $E_{\textrm{w}}$ of the interacting ferromagnet bands; and the interaction parameters (IA) necessary to fit the DFT data of the corresponding structure with the $p_z$-$d$ model Hamiltonian $\mathcal{H}$, Eq. (\ref{eq:Exchange_Hamiltonian}).
}
\end{table*}

Figure \ref{fig:Spin_bands_Co_2hBN_Gr}(a) shows the spin-polarized band structure of graphene/hBN/Co for two layers of hBN. 
In the band structure, now, the \textit{spin-up} graphene Dirac states are no longer lying lower in energy than the \textit{spin-down} ones, leading to a reversal of the sign of the exchange parameters $\lambda_{\textrm{ex}}^{\textrm{A}}$ and $\lambda_{\textrm{ex}}^{\textrm{B}}$. 
The band structure shows that the doping level decreases by roughly $80$~meV, and the hybridization with the $d$ band with energy $E_{\textrm{v}}$, coming from the top Co layer, is strongly enhanced, in contrast to the one-hBN-layer case.
The perfect fit to the $p_z$-$d$ model is also seen in Fig.~\ref{fig:Spin_bands_Co_2hBN_Gr}(b); the obtained parameters are given in Table \ref{tab:summary} for Co as the ferromagnet and two layers of hBN. 
In Fig.~\ref{fig:Spin_bands_Co_2hBN_Gr}(c) we can see that the band splitting at the K point of the conduction (valence) bands is smaller (larger) than in the one-hBN-layer case. In addition, the proximity-induced gap nearly doubles, and the hybridization to the Co $d_{z^2}(3)$ state is much stronger.

The inset in Fig. \ref{fig:Spin_bands_Co_2hBN_Gr}(b) shows that for the case of two hBN layers, carbon atom $\textrm{C}_{\textrm{B}}$ (now at the top site above Co) has a direct connection to the Co atom in the top position via a nitrogen atom and a boron atom of the two individual hBN layers. 
Localized at this Co atom in the top position, there is some density with $d_{z^2}$ character (resulting in the band with energy $E_{\textrm{v}}$) which can propagate through this direct path and polarize carbon atom $\textrm{C}_{\textrm{B}}$. 
This hybridization is described by the parameter $\textrm{v}_{\uparrow}^{\textrm{B}}$, shifting the corresponding bands in energy and leading to the opening of a hybridization gap in the band structure. 
The vertical stacking of the atoms facilitates the hybridization of the carbon $p_z$ states with Co $d$ states.  
For the case of one hBN layer there is no direct path connecting Co atoms in the top position and carbon atoms $\textrm{C}_{\textrm{B}}$, and thus the hybridization is suppressed. 
Again, by employing our $p_z$ model directly at the K point, we can extract parameters which correspond to the pure splittings of the Dirac bands at the K point, as given in Fig. \ref{fig:Spin_bands_Co_2hBN_Gr}(c). 

\begin{table}[htb]
\begin{ruledtabular}
\begin{tabular}{lccccc}
FM&
hBN (layers)&
\textrm{$E_{\textrm{D}}$~(meV)}&
\textrm{$\Delta $~(meV)}&
\textrm{$\lambda_{\textrm{ex}}^{\textrm{A}}$~(meV)} &
\textrm{$\lambda_{\textrm{ex}}^{\textrm{B}}$~(meV)} \\
\hline
Co & 1 & -433.10 &  19.25 & -3.14 & 8.59\\
 & 2 & -348.03 &  36.44 & 0.097 & -9.81\\
 & 3 &-301.10 & 38.96 & -0.005 & 0.018\\
Ni & 1 & -527.89 & 22.86 & -1.40 & 7.78\\
   & 2 & -434.82 & 42.04 & 0.068 & -3.38\\
   & 3 & -361.57 & 40.57 & -0.005 & 0.017\\

\end{tabular}
\end{ruledtabular}
\caption{\label{tab:summary_pz}Summary of the most relevant parameters for all relevant structures ($a=2.489$~$\textrm{\AA}$ and $U = 0$~eV) for the different ferromagnets (FM) Co and Ni for one to three layers of hBN: proximity gap $\Delta$, Dirac point energy $E_{\textrm{D}}$, and the exchange parameters $\lambda_{\textrm{ex}}^{\textrm{A}}$ and $\lambda_{\textrm{ex}}^{\textrm{B}}$ necessary to fit the DFT data of the corresponding structure with the minimal $p_z$ model Hamiltonian, Eq. (\ref{eq:pex_Hamiltonian}).}
\end{table}

The values of $\tilde{\lambda}_{\textrm{ex}}^{\textrm{B}}$ and $\lambda_{\textrm{ex}}^{\textrm{B}}$, obtained from the two models, are given in Tables \ref{tab:summary} and \ref{tab:summary_pz}. 
They deviate by a factor of 20, which comes from the fact that the minimal $p_z$ model describes dressed exchange
parameters, whereas the $p_z$-$d$ model describes the bare exchange parameters. The dressed parameters 
contain both the interlayer exchange and spin-selective hybridization of the $p_z$ and $d$ orbitals. The bare
exchange couplings $\tilde{\lambda}_{\rm ex}$ are much weaker than in the single-hBN-layer case, by an order of magnitude.
However, the dressed coupling $\lambda_{\rm ex}^{\rm B}$ stays at a similar magnitude (the sign changes). The reason is that the
valence-band spin splitting is dominated by the anticrossing of $d_{z^2}(3)$ and $p_z(\textrm{C}_{\textrm{B}})$ orbitals, affecting
only the spin-up component. The spin-down valence-band is not affected. As a result, the proximity spin splitting
is, in this case, caused by shifting the spin-up band relative to its spin-down counterpart, by the spin-selective
hybridization. {\it This mechanism of proximity exchange can lead to a giant enhancement of the 
proximity spin splittings which can be tailored by the electric field.}

\subsubsection{Additional considerations}
\label{subsubsec:3hBN_Co}

Here we address some outstanding questions related to our above analysis. How do additional insulating layers perform? Can we tune the doping level by an external electric field? Is the proximity exchange affected? Are the band splittings we see representative of a thick ferromagnetic substrate (are three Co layers enough)? How is the proximity effect affected
by the Hubbard $U$, which shifts the $d$ orbital levels?

In the following we consider only the dressed band splittings $\lambda_{\rm ex}$, obtained with the minimal $p_z$ model
directly at the K point. Bare splittings are barely affected by electric fields, and their behavior with respect to the number
of layers is that of a damped oscillator.

\begin{figure}[htb] 
 \centering 
\includegraphics[width=0.49\textwidth]{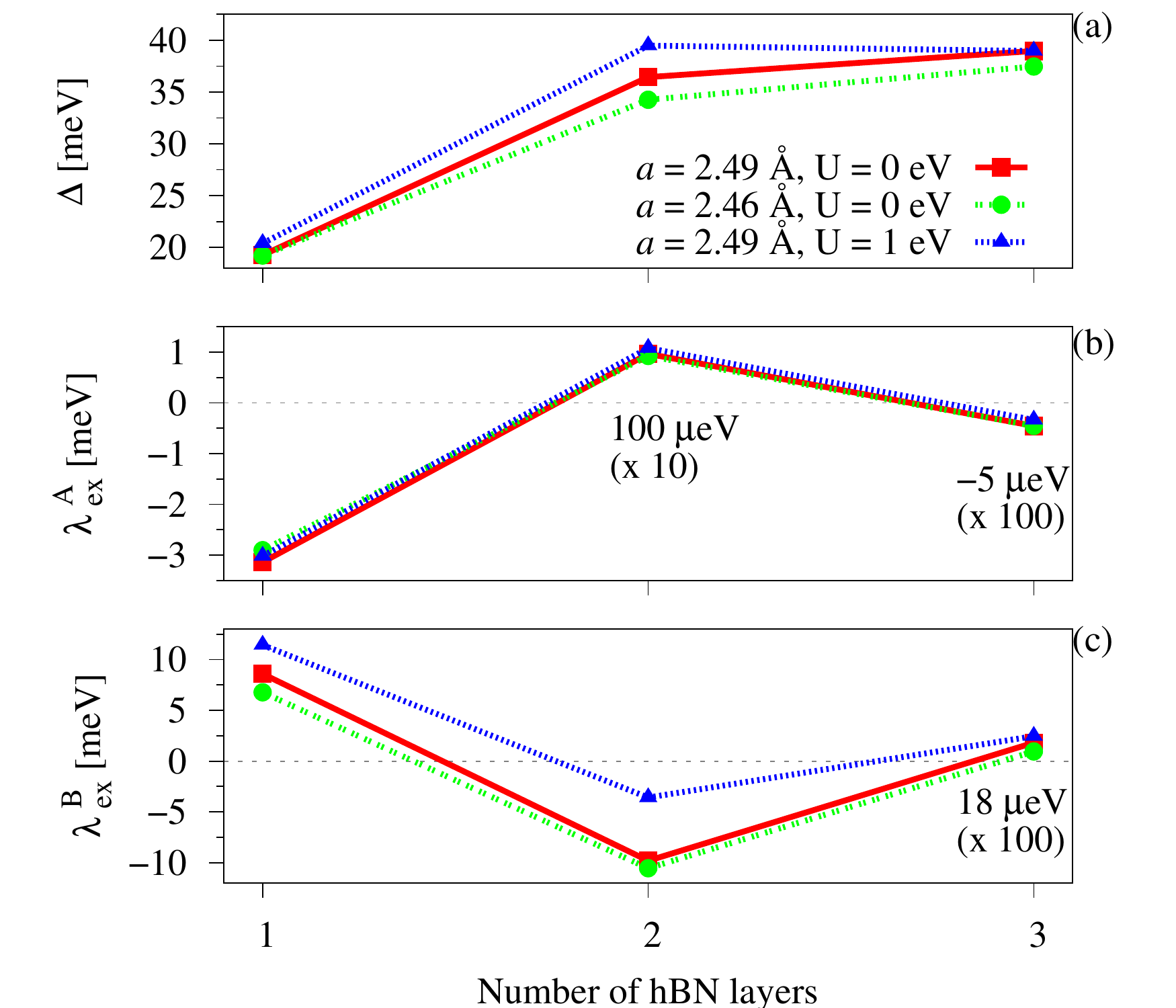}
 \caption{\label{fig:hBN_layers_Co}(Color online) Influence of the number of hBN layers on the proximity-induced parameters for the graphene/hBN/Co structure, using the $p_z$ model at the K point. Dependence of (a) the proximity gap $\Delta$, (b) the exchange parameters $\lambda_{\textrm{ex}}^{\textrm{A}}$, and (c) 
$\lambda_{\textrm{ex}}^{\textrm{B}}$ on the number of hBN layers for different lattice constants or an additional Hubbard parameter of $U = 1.0$~eV. 
Parameter values for two (three) layers of hBN were increased by a factor of 10 (100) for better visualization as indicated. }
 \end{figure}

\paragraph{Dependence on the number of hBN layers.}
Figure \ref{fig:hBN_layers_Co} shows the dependence of the proximity gap $\Delta$ and the two exchange parameters $\lambda_{\textrm{ex}}^{\textrm{A}}$ and $\lambda_{\textrm{ex}}^{\textrm{B}}$ on the number of hBN layers between Co and graphene. 
We can see that the exchange parameters change sign after an additional insulating layer is added. 
The proximity gap $\Delta$ nearly doubles for two layers of hBN and stays essentially unchanged when a 
third layer is added since the local environment of graphene does not change anymore. 
The parameters obtained with the minimal proximity exchange model are summarized in 
Table \ref{tab:summary_pz} for $a=2.489$~$\textrm{\AA}$ and Hubbard $U = 0$.
For four layers of hBN, again, the parameters change sign, but they are even smaller than for three layers of hBN and thus are not included here. The parameters $\lambda_{\textrm{ex}}^{\textrm{A}}$ and $\lambda_{\textrm{ex}}^{\textrm{B}}$ are already in the $\mu$eV regime for three hBN layers, which is a result of the strong barrier. The two-layer case is
special for the case of $\lambda_{\rm ex}^{\rm B}$ since this exchange parameter has a magnitude similar to that of the
single-layer case due to the strong hybridization with the $d$ orbitals close to the K point. 
We note that the distances for the three-layer case are similar to those for the two-layer case. 
We only have one additional distance between the two hBN layers directly below graphene, which was relaxed to $d_{\textrm{hBN/hBN}} = 3.088$~${\textrm{\AA}}$.

As we have already seen, the bands of hBN are also spin split. To get the magnitude of the exchange splitting of the individual hBN layers, we look at the graphene/hBN/Co structure with three layers of hBN. 
In the band structure we can identify the highest- (lowest-) lying valence (conduction) bands, which are spin split, of the three individual layers, like in Fig. \ref{fig:Effects}(a). From that, we extract the band splittings of conduction $\Delta E_{\textrm{cond}}$ and valence $\Delta E_{\textrm{val}}$ bands of the individual hBN layers at the K point. 
The spin-up bands of hBN are always lying lower in energy than the spin-down ones at the K point. 
 \begin{figure}[htb] 
 \centering 
\includegraphics[width=0.49\textwidth]{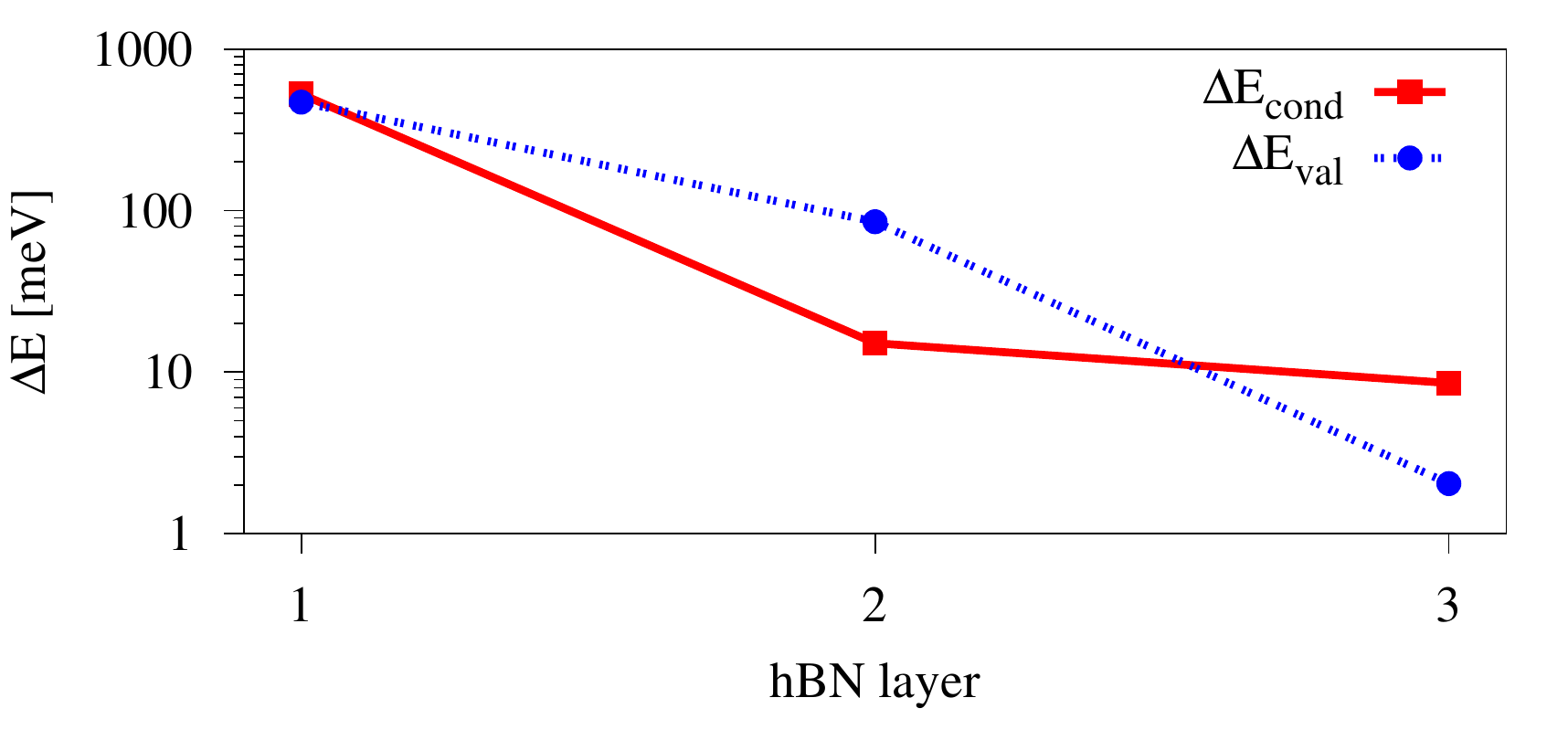}
 \caption{\label{fig:splittings_hBN_cobalt}(Color online) Conduction $\Delta E_{\textrm{cond}}$ and valence $\Delta E_{\textrm{val}}$ band splittings of the three individual hBN layers at the K point. Values are obtained by identifying the spin-split hBN conduction and valence bands of the three individual layers in the band structure of the graphene/hBN/Co heterostructure for three layers of hBN.}
 \end{figure}
Due to the spin splitting of the bands, hBN can additionally act as a spin filter for tunneling electrons, as reported in Ref. \onlinecite{Kamalakar2016}.
In Fig. \ref{fig:splittings_hBN_cobalt} we show the valence- and conduction-band splittings at the K point of the three hBN layers. We find that the exchange splitting of the first hBN layer (closest to the Co surface) is roughly $0.5$~eV. The splittings of the second and third layers are exponentially suppressed, but the third layer still exhibits
proximity exchange of more than 1 meV (10 meV for the conduction band).

\paragraph{Lattice-constant effects.}
Since we have artificially set the lattice constant for all the (well-lattice-matched) materials to be the same value, we now
consider its effect on the proximity structure. We use the graphene constant $a = 2.46$~${\textrm{\AA}}$ by simply changing the in-plane lattice constant of the slab to this value without changing the vertical distances between the layers, which should be more favorable for the description of the graphene dispersion. 
The results in this case do not deviate much from the case with our adopted $a = 2.489$~${\textrm{\AA}}$, as can be seen in Fig. \ref{fig:hBN_layers_Co}, but the Fermi velocity for $a = 2.46$~${\textrm{\AA}}$ and one hBN layer is 
$v_{\textrm{F}} = 0.827\times 10^6$~m/s, corresponding to a larger nearest-neighbor hopping parameter of $t=2.56$~eV.

\paragraph{Hubbard U.}
Since the exact position of the $d$ bands is crucial  to see the giant proximity exchange in the case of two hBN
layers, we consider what happens when we apply a Hubbard $U$ parameter to the calculation and shift the
$d$-orbital levels. From recent studies of graphene on copper \cite{Frank2016} we know that the copper bands have to be shifted down in energy by $U=1.0$~eV to match the measured band structure from angle-resolved photoemission spectroscopy experiments. From other DFT studies \cite{Nakamura2006, Aryasetiawan2006, Svane1990, Juhin2010, Forti2012}, mainly on metal oxides, it is not possible to get a unique value for $U$.
Thus we apply $U = 1.0$~eV as a generic representative. The results are shown in Fig. \ref{fig:hBN_layers_Co}.
We can see that the parameters $\Delta$ and $\lambda_{\textrm{ex}}^{\textrm{A}}$ stay almost unchanged, as 
they are not affected by the strong coupling with $d$ orbitals. However, $\lambda_{\textrm{ex}}^{\textrm{B}}$, representing the valence Dirac band splitting is strongly affected, especially in the case of two hBN layers (it is not affected for
three layers). By applying the Hubbard $U$, we shift the band with energy $E_{\textrm{v}}$ in Fig. \ref{fig:Spin_bands_Co_2hBN_Gr} down, away from the Dirac states, so the splitting at the K point decreases. 
The energetic position of the $d$ bands with respect to the Dirac bands strongly influences the pure band splittings at the K point if the hybridization is large. In the absence of experimental guidance into the exact relative position of $d$
levels in our system, we can thus only predict the general trends and rough magnitudes for the valence proximity
splitting. If the $d$ bands are indeed close to the Dirac point, their influence will be giant, and one can expect ramifications
in spin tunneling and spin injection. 

\paragraph{Electric field effects.}
 \begin{figure}[htb] 
 \centering 
 \includegraphics[width=0.48\textwidth]{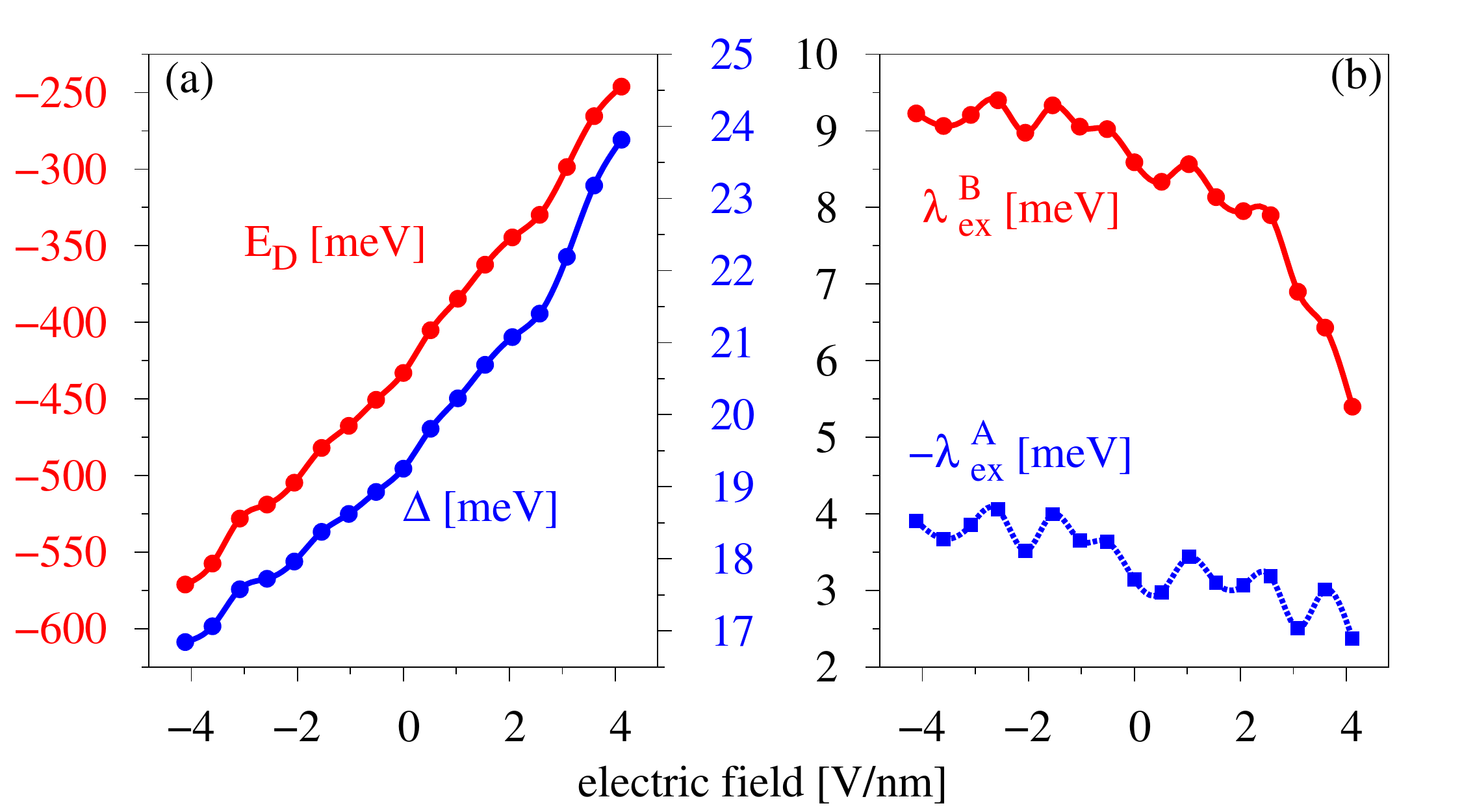}
 \caption{\label{fig:Efield_Co}(Color online) Influence of the electric field on the proximity-induced parameters for the graphene/hBN/Co structure for one hBN layer, using the minimal $p_z$ model at the K point. Dependence of the (a) Dirac energy $E_{\textrm{D}}$ and the proximity gap $\Delta$ and (b) the exchange parameters $\lambda_{\textrm{ex}}^{\textrm{A}}$ and $\lambda_{\textrm{ex}}^{\textrm{B}}$ on the applied transverse electric field.}  
 \end{figure} 
 \begin{figure}[htb] 
 \centering 
 \includegraphics[width=0.48\textwidth]{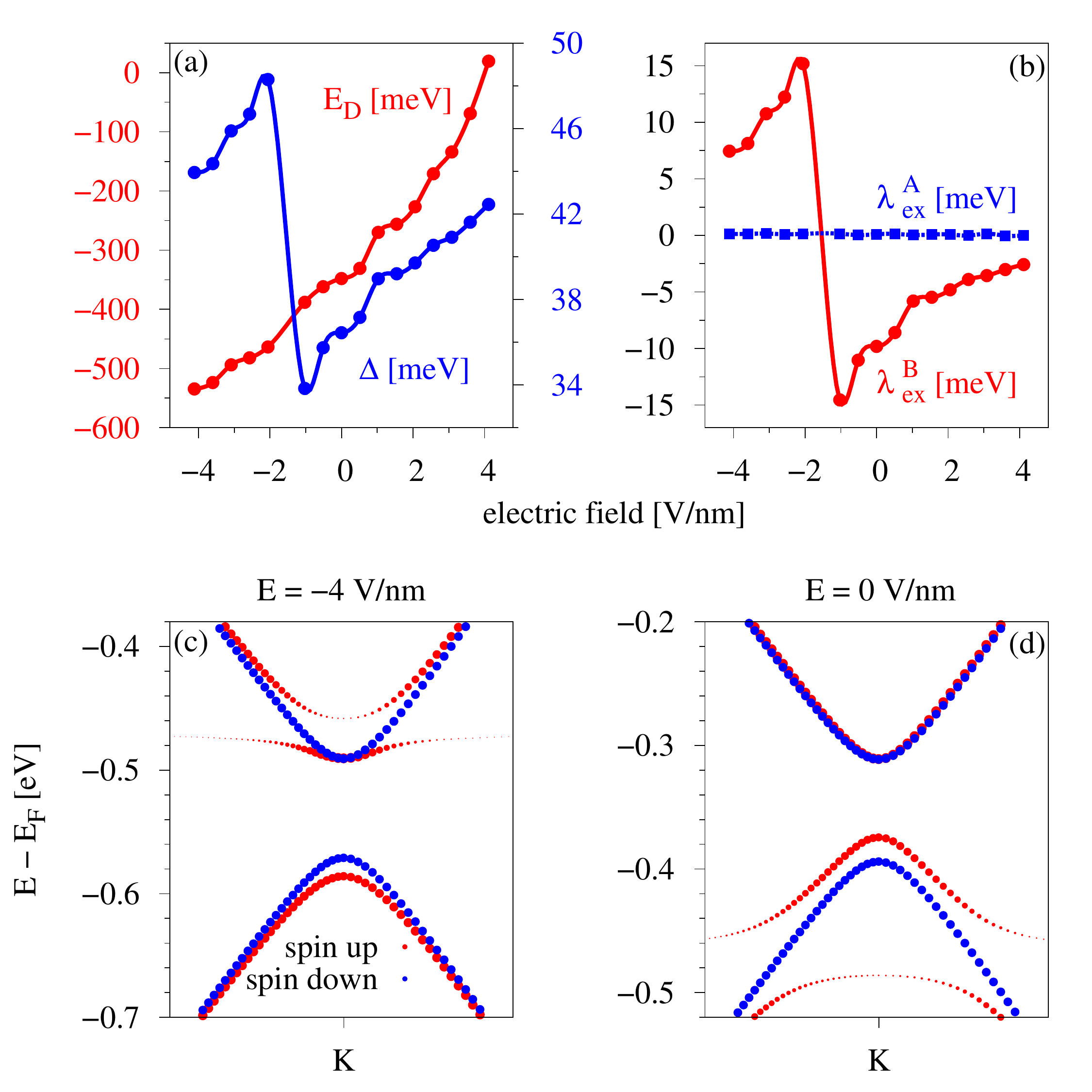}
 \caption{\label{fig:Efield_Co_2hBN}(Color online) Influence of the electric field on the proximity-induced parameters for the graphene/hBN/Co structure for two hBN layers, using the minimal $p_z$ model at the K point. Dependence of the (a) Dirac energy $E_{\textrm{D}}$ and the proximity gap $\Delta$ and (b) the exchange parameters $\lambda_{\textrm{ex}}^{\textrm{A}}$ and $\lambda_{\textrm{ex}}^{\textrm{B}}$ on the applied transverse electric field. (c) and (d) The calculated spin-resolved band structure projected on the graphene states in the vicinity of the Dirac point for two different field strengths, illustrating the reversal of the valence spin states at the K point.}  
 \end{figure} 

Figure \ref{fig:Efield_Co} shows the influence of a transverse electric field on the proximity parameters $\Delta$, $\lambda_{\textrm{ex}}^{\textrm{A}}$, and $\lambda_{\textrm{ex}}^{\textrm{B}}$ and the Dirac point energy $E_{\textrm{D}}$ for one layer of hBN. 
We model our electric field with a saw-like potential oriented perpendicular to the slab structure. 
A positive field points from cobalt towards graphene and depletes its conduction electrons (lowers the magnitude of $E_{\rm D}$).

We can see that $E_{\textrm{D}}$ and $\Delta$ show the same trend with electric field.
In general, by increasing the electric field the doping level decreases; that is, one just shifts the Fermi level with respect to the Dirac bands. 
The proximity gap $\Delta$ also increases with increasing electric field, reflecting the charge transfer away from graphene. The continuous shift of the doping level with the applied electric field allows us to shift the Fermi level to the desired position.
The general trend of the proximity parameters is that both tend to  decrease with increasing electric field.
For moderate field strengths of $\pm 2$~V/nm, the parameters and thus the band splittings at the K point are almost unaffected. 

Most interesting is the two-hBN-layer case since here the valence-band splitting is strongly affected by hybridization with a $d$ level. 
By applying an electric field, we can tune the energetic position of the Dirac point with respect to the $d$ levels, which should also strongly affect the spin splitting of the graphene Dirac bands. 
In Fig. \ref{fig:Efield_Co_2hBN} we show the influence of the electric field on the proximity parameters for two layers of hBN.
We can see that the Dirac point energy $E_{\textrm{D}}$ increases with electric field, as for the monolayer hBN case. 
The proximity parameter $\lambda_{\textrm{ex}}^{\textrm{A}}$ stays constant in magnitude around $100~\mu$eV.
In Figs. \ref{fig:Efield_Co_2hBN}(c) and \ref{fig:Efield_Co_2hBN}(d), we show the calculated spin-resolved band structure of the graphene/hBN/Co heterostructure for two hBN layers, projected on the graphene states in the vicinity of the Dirac point for different field strengths. 
The spin-up graphene valence band at the K point is lying lower in energy than the spin-down one for $E = -4$~V/nm and vice versa for $E = 0$~V/nm. 
Therefore the parameter $\lambda_{\textrm{ex}}^{\textrm{B}}$ is positive (negative) for fields smaller (larger) than $-1.5$~V/nm [see Fig. \ref{fig:Efield_Co_2hBN}(b)]. 
The {\it crossover} happens at about $-1.5$~V/nm. 
The reason for these reversing spin states is the resonant $d$ level. 
At a certain energetic configuration between the Dirac point and the $d$ level, adjusted by the external electric field, the hybridization of the $d$ level with graphene valence $p_z$ states leads to the change of the sign of the spin splitting parameters.
{\it This allows us to control the sign of the injected spin by applying an electric field}, shifting the Dirac bands through the resonant $d$ level. 
(Of course, this effect can only be observed if the $d$ bands are indeed close to the Dirac point. DFT calculations
can provide, at most, indications of this occurring, due to the insufficient treatment of correlations that are important for 
$d$ orbitals of transition metals.) 
Also, the proximity gap $\Delta$ jumps in magnitude at the same field strength, roughly $-1.5$~V/nm, since the parameters $\Delta$ and $\lambda_{\textrm{ex}}^{\textrm{B}}$ of the $p_z$ model are connected. Apart from the jump, the gap parameter increases with increasing field strength.

\begin{figure}[!htb] 
 \centering 
 \includegraphics[width=0.48\textwidth]{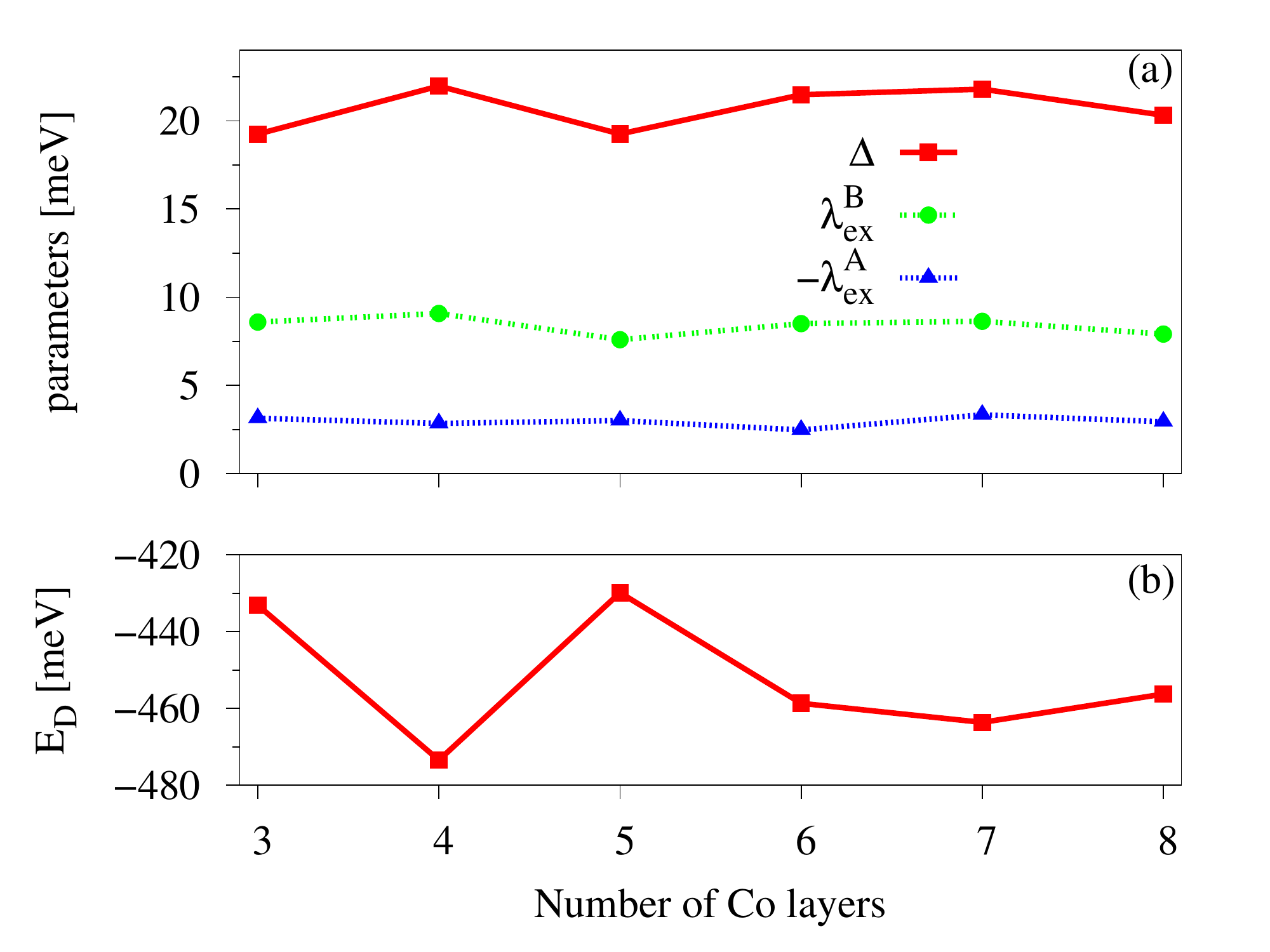}
 \caption{\label{fig:Co_layers}(Color online) Influence of the number of Co layers on the band structure for the graphene/hBN/Co system for one hBN layer, using the minimal $p_z$ model at the K point. Dependence of (a) the proximity gap $\Delta$ and the exchange parameters $\lambda_{\textrm{ex}}^{\textrm{A}}$ and $\lambda_{\textrm{ex}}^{\textrm{B}}$, as well as (b) the Dirac energy $E_{\textrm{D}}$, on the number of Co layers.}  
 \end{figure}

\paragraph{Additional cobalt layers.}
Finally, we analyze the influence of additional Co layers on the band structure (Fig. \ref{fig:Co_layers}). As we increase the number of Co layers, more $d$ bands are introduced into the dispersion.
Consequently, in the vicinity of the K point in Fig. \ref{fig:Effects}(a) graphene states can be disturbed by these additional Co bands. 

We can see that the band splittings of the graphene Dirac states at the K point do not get influenced much by additional layers, since the parameters $\lambda_{\textrm{ex}}^{\textrm{A}}$ and $\lambda_{\textrm{ex}}^{\textrm{B}}$ stay almost constant, but the Dirac energy, which is our measure for the doping level, saturates only after six Co layers are present. 
We conclude that three Co layers suffice to obtain representative proximity parameters, and six Co layers are needed to 
fix the relative positioning of the bands.  

\subsection{Graphene/hBN/nickel}
\subsubsection{Structure}
\begin{figure}[htb] 
 \centering 
 \includegraphics[width=0.48\textwidth]{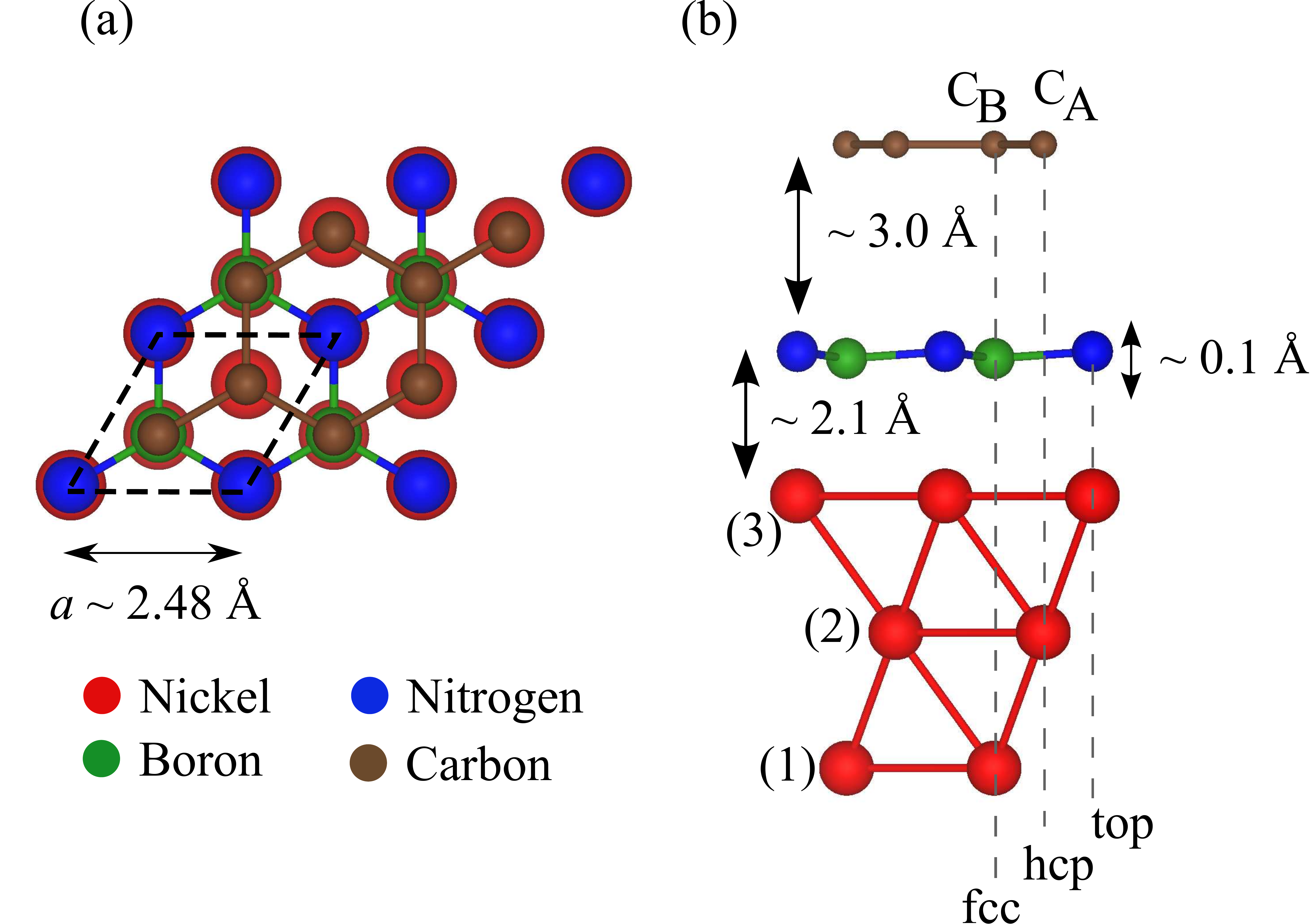}
 \caption{\label{fig:Structure_Ni}(Color online) Structure of graphene/hBN/Ni, with labels for the different atoms.
 (a) Top view of the structure, with one unit cell emphasized by the dashed line. (b) Side view with stacking configuration: $\textrm{C}_{\textrm{B}}$ is over boron, and $\textrm{C}_{\textrm{A}}$ is over the hBN ring. Nitrogen is at the top site above Ni, and boron is above the fcc site of Ni. The indicated distances are measured between graphene/Ni and the nitrogen atom of hBN since the hBN layer is slightly corrugated by $\Delta z = 0.101$~${\textrm{\AA}}$, with the boron atom  closer to the Ni surface. Numbers in parenthesis indicate the Ni layer.}  
 \end{figure}

We now use Ni as the ferromagnet like in the approach with Co. 
Nickel crystallizes in a fcc lattice and has a magnetic moment of about $0.6~\mu_B$, smaller than the one for hcp cobalt, which is $1.7~\mu_B$ \cite{kittel2004introduction}. Thus we expect the effects of proximity-induced magnetism to be smaller for the Ni substrate.
In order to stack a hexagonal lattice on top of it, we need to consider the (111) plane. The lattice constant of Ni \cite{kittel2004introduction} is $ a = 3.524$~${\textrm{\AA}}$, and thus the lattice constant of the quasi-hexagonal lattice of the (111) plane is $\frac{1}{2}\sqrt{2}a = 2.492$~${\textrm{\AA}}$. As a result, the (111) plane of Ni is suitable for making heterostructures with graphene; the lattice mismatch is small.
We fix an effective average lattice constant of $a = 2.48$~${\textrm{\AA}}$ for the systems with Ni.
In this case, the lowest-energy configuration is when nitrogen atoms are at the top sites above Ni and boron atoms are at fcc sites above Ni. Carbon atoms sit on top of boron atoms and at the hollow sites, above the center of a hexagonal ring of hBN (see Fig. \ref{fig:Structure_Ni}), in agreement with previous DFT studies \cite{Bokdam2013, Bokdam2014, Giovannetti2007}.

After relaxation of atomic positions we obtained layer distances of $d_{\textrm{Ni/hBN}} = 2.105$~${\textrm{\AA}}$ between Ni and hBN and $d_{\textrm{hBN/Gr}} = 3.015$~${\textrm{\AA}}$ between hBN and graphene (measured between C/Ni and N atoms, respectively, since the hBN layer is corrugated). \\
The layer distances of this minimum-energy configuration are in agreement with Refs. \onlinecite{Giovannetti2007, Vanin2010, Bokdam2014}, which report $d_{\textrm{hBN/Gr}} = 3.22 - 3.40$~${\textrm{\AA}}$ and $d_{\textrm{Ni/hBN}} = 1.96 - 2.12$~${\textrm{\AA}}$. 
Again, the hBN-layer is not flat anymore but slightly corrugated by $0.101$~${\textrm{\AA}}$, in agreement with Refs. \onlinecite{Bokdam2014, Grad2003}.
For hBN we use an AA$'$ stacking (B over N, N over B), which is the energetically favorable one, with distances between the layers in the range of $d_{\textrm{hBN/hBN}} = 2.99 - 3.08$~${\textrm{\AA}}$ (details are given in sections \ref{subsubsec:2hBN_Ni} and \ref{subsubsec:3hBN_Ni}).

\begin{figure*}[htb] 
 \centering 
 \includegraphics[width=0.96\textwidth]{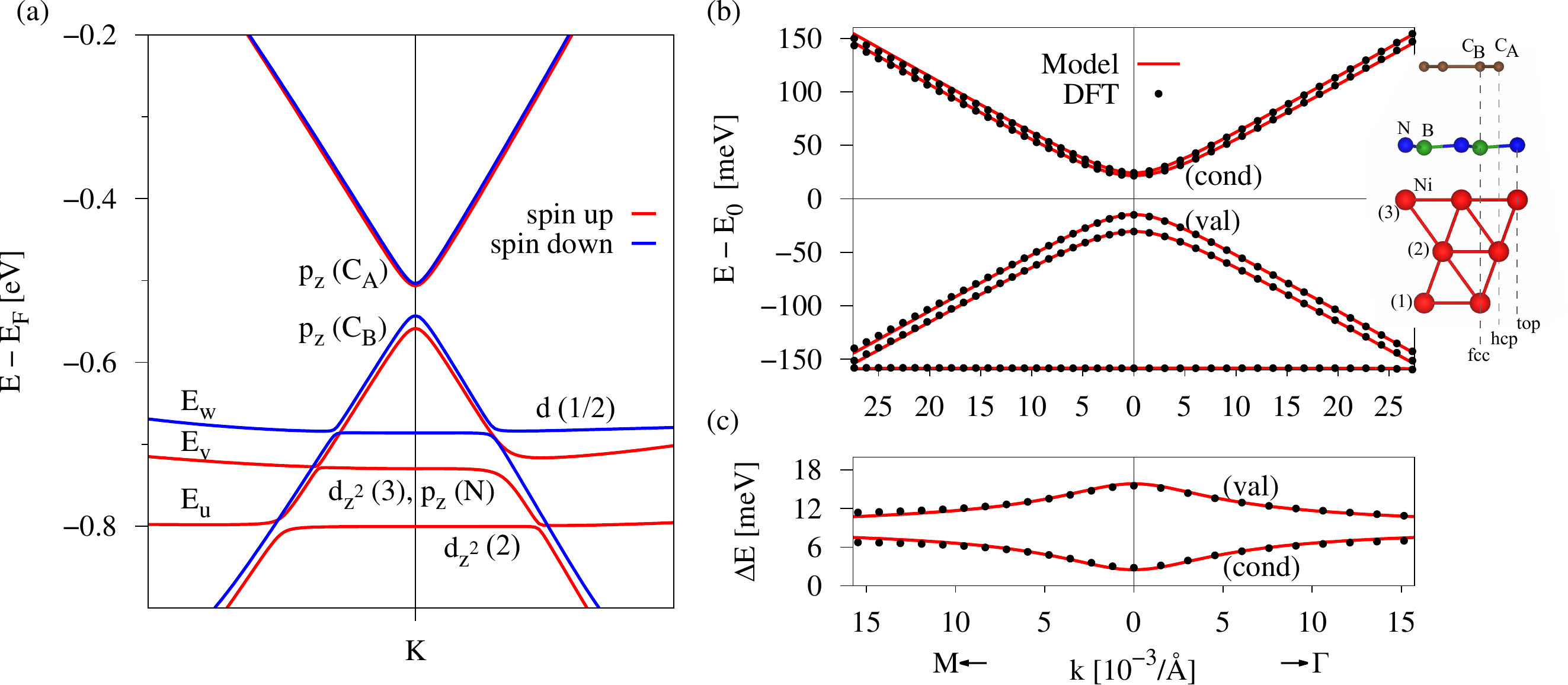}
 \caption{\label{fig:Spin_bands_Ni_1hBN_Gr}(Color online) Spin-polarized band structure of the graphene/hBN/Ni heterostructures for one layer of hBN. (a) Band structure in the vicinity of the Dirac point with labels for the main orbital contributions. Labels $E_{j}$, $j = \textrm{u,v,w}$, are the energy bands, which correspond to the Ni $d$ states used to fit the $p_z$-$d$ model Hamiltonian in Eq. (\ref{eq:Exchange_Hamiltonian}). (b) The fit to the $p_z$-$d$ model with a side view of the structure. First-principles data (dotted lines) are well reproduced by the model (solid lines). (c) The corresponding splittings of the valence (val) and conduction (cond) Dirac states of graphene. The fit parameters are $E_{0} = -527.98$~meV, $\tilde{\Delta} = 22.98$~meV, $\tilde{\lambda}_{\textrm{ex}}^{\textrm{A}} = -1.25$~meV, $\tilde{\lambda}_{\textrm{ex}}^{\textrm{B}} = 8.17$~meV, $E_{\textrm{u}} = -272.58$~meV, $E_{\textrm{v}} = -201.27$~meV, $E_{\textrm{w}} = -158.17$~meV, $\textrm{v}_{\uparrow}^{\textrm{B}}= 10.15$~meV, $\textrm{w}_{\downarrow}^{\textrm{B}}= 4.19$~meV.
The Fermi velocity to match the slope away from the K point is $v_{\textrm{F}} = 0.81\times 10^6$ m/s. The fit parameters are again obtained in the same way as for the Co case.}  
 \end{figure*}

\subsubsection{One hBN layer}
Figure \ref{fig:Spin_bands_Ni_1hBN_Gr}(a) shows the calculated spin-polarized band structure of the graphene/hBN/Ni heterostructure for one layer of hBN.
The graphene Dirac states for \textit{spin up} are lying lower in energy than the \textit{spin-down} ones, as in the Co case.
Comparing Ni and Co, we notice that the Dirac point energy $E_{\textrm{D}}$ for Ni is about $100$~meV lower than for Co, but the proximity-induced band splittings are smaller, as expected due to the smaller magnetic moment of Ni. 
In general the band structures are quite similar, with the difference being that Ni $d$ states do not influence the Dirac states as much as Co does.
Additionally, we notice that the \textit{spin-up} $d$ bands are formed by the same orbitals as for Co, while the \textit{spin-down} $d$ band near the K point is formed by different orbitals [mixture of $d$ orbitals of Ni layers (1/2) except $d_{z^2}$] due to the different lattices of Ni and Co. 
Most of all, we notice that there is no $d$ band crossing the conduction Dirac states in the relevant energy and $k$ region.

The fit to the $p_z$-$d$ model is shown by solid lines to the DFT data in Fig.~\ref{fig:Spin_bands_Ni_1hBN_Gr}(b).
We see that the $p_z$-$d$ model Hamiltonian, Eq. (\ref{eq:Exchange_Hamiltonian}), describes our first-principles results very well with the fit parameters given in Table \ref{tab:summary}. Like in the Co case, the gap in the dispersion is roughly $40$~meV, and the band splittings are of the order of $10$~meV. 
We additionally employ our  minimal model to extract the effective band spin splittings (see Table \ref{tab:summary_pz}). 
Due to the weak hybridization with $d$ orbitals, the minimal model parameters are very close to the parameters
of the $p_z$-$d$ model.

 \begin{figure*}[!htb] 
 \centering 
 \includegraphics[width=0.96\textwidth]{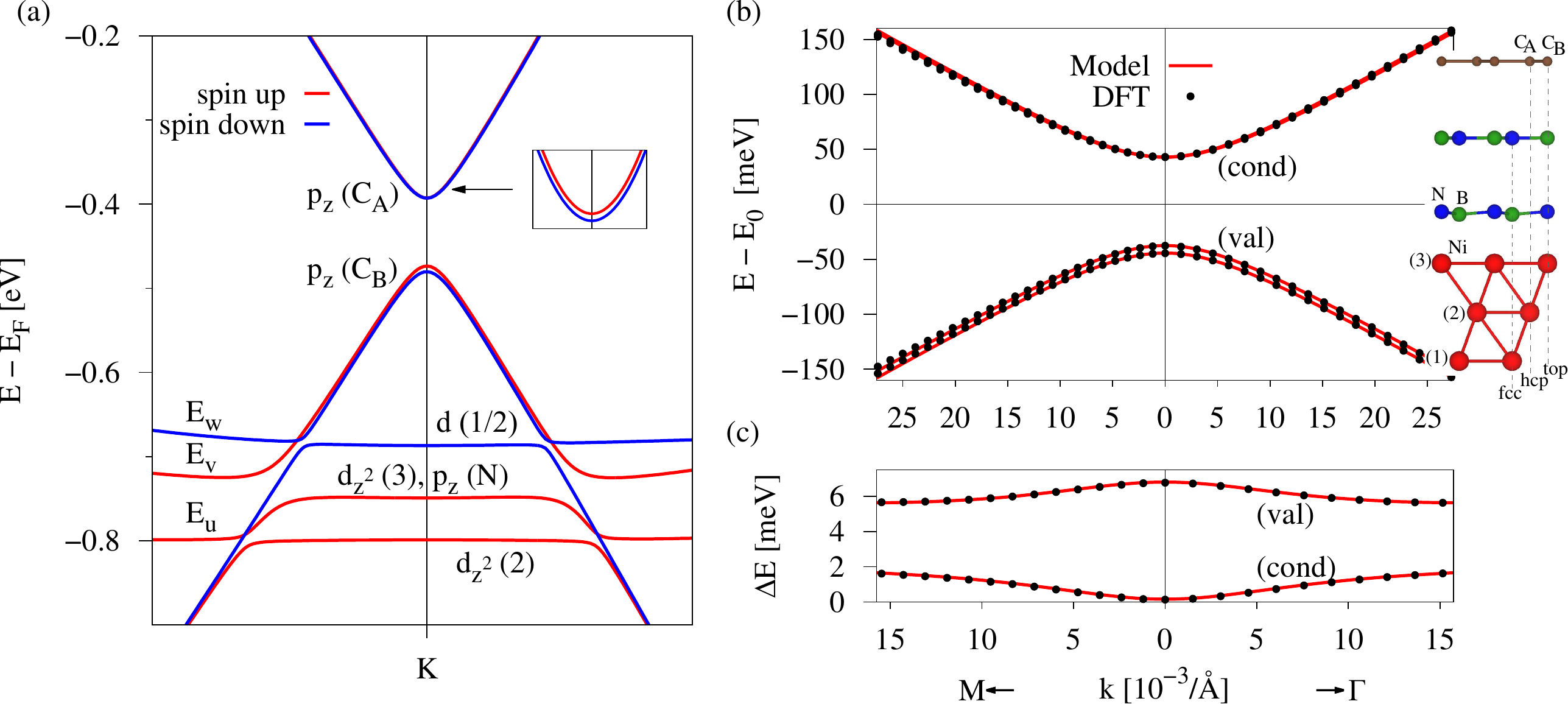}
 \caption{\label{fig:Spin_bands_Ni_2hBN_Gr}(Color online) Spin -polarized band structure of graphene/hBN/Ni heterostructures for two layers of hBN (AA$'$ stacking). (a) Band structure in the vicinity of the Dirac point with labels for the main orbital contributions. The inset shows close-up of the conduction Dirac states to visualize the reversal of the spin states. (b) The fit to the $p_z$-$d$ model with a side view of the structure for two layers of hBN. First-principles data (dotted lines) are well reproduced by the $p_z$-$d$ model (solid lines). (c) The corresponding splittings of the valence and conduction Dirac states. The fit parameters are $E_{0} = -435.76$~meV, $\tilde{\Delta} = 42.88$~meV, $\tilde{\lambda}_{\textrm{ex}}^{\textrm{A}} = 0.080$~meV, $\tilde{\lambda}_{\textrm{ex}}^{\textrm{B}} = -1.44$~meV, $E_{\textrm{u}} = -363.36$~meV, $E_{\textrm{v}} = -309.27$~meV, $E_{\textrm{w}} = -251.05$~meV, $\textrm{v}_{\uparrow}^{\textrm{B}}= 32.67$~meV.
The Fermi velocity to match the slope away from the K point is $v_{\textrm{F}} = 0.824\times 10^6$ m/s. All other parameters are zero for the same fitting range as for the one-layer case.}  
 \end{figure*}
 
\subsubsection{Two hBN layers}
\label{subsubsec:2hBN_Ni}
Figure \ref{fig:Spin_bands_Ni_2hBN_Gr} shows the calculated band structure and the fit to the $p_z$-$d$ model in the case of two layers of hBN and three layers of Ni.
Again, the positions of carbon $\textrm{C}_{\textrm{A}}$ did not change with respect to the hBN layers, while the position of $\textrm{C}_{\textrm{B}}$ was changed to be on top of the uppermost boron atom. 
The layer distance between the two hBN layers was relaxed to
$d_{\textrm{hBN/hBN}} = 2.995$~${\textrm{\AA}}$ and the distance between the uppermost hBN layer and graphene was $d_{\textrm{hBN/Gr}} = 3.110$~${\textrm{\AA}}$ in the two-layer case. The corrugation of the lower hBN layer and the distance between hBN and Ni did not change. The inset in Fig. \ref{fig:Spin_bands_Ni_2hBN_Gr}(b) shows the geometry for two layers of hBN.
Figure \ref{fig:Spin_bands_Ni_2hBN_Gr}(a) shows the spin-polarized band structure of graphene/hBN/Ni for two layers of hBN. 
The \textit{spin-up} graphene Dirac states are again no longer lying lower in energy than the \textit{spin-down} ones, leading to the reversal of the sign of the exchange parameters, just as for Co. 
The fit parameters for the $p_z$-$d$ model are given in Table \ref{tab:summary}.
The fit to the $p_z$-$d$ model is shown in Fig.~\ref{fig:Spin_bands_Ni_2hBN_Gr}(b).

We can see that the band splittings for both the conduction and valence Dirac states are smaller than in the single-hBN-layer case, as expected due to the additional insulating layer, while 
the proximity-induced gap $\Delta$ nearly doubles, and the hybridization to the Ni $d_{z^2}(3)$ state is much larger. From the geometry in Fig.~\ref{fig:Spin_bands_Ni_2hBN_Gr}(b), we can again notice that carbon $\textrm{C}_{\textrm{B}}$ orbitals can couple to $d$ orbitals of Ni in the top position via a nitrogen atom and a boron atom of the two individual hBN layers, which is responsible for the strong hybridization with the $d$ band with energy $E_{\textrm{v}}$. This hybridization drives the strong proximity exchange in the valence band of graphene.

By employing our  minimal model directly at the K point we extract the effective exchange parameters corresponding to the values of the splittings in Fig. \ref{fig:Spin_bands_Ni_2hBN_Gr}(c). The parameters are
summarized in Table \ref{tab:summary_pz}. If we compare $\tilde{\lambda}_{\textrm{ex}}^{\textrm{B}}$ and $\lambda_{\textrm{ex}}^{\textrm{B}}$, we see that they are of similar magnitudes (unlike for the Co case)
since the $d$ band with energy $E_{\textrm{v}}$ is relatively far away from the Dirac point energy, so that 
the hybridization effects on the band splittings at the K point are similar in monolayer and bilayer hBN structures. There
is no resonant $d$ level as in the Co case. 

 \subsubsection{Additional considerations}
 \label{subsubsec:3hBN_Ni}
In the following, we consider effective band splittings directly at the K point, which correspond to the exchange couplings in the minimal model. 

\paragraph{Dependence on the number of hBN layers.}

\begin{figure}[htb] 
 \centering 
 \includegraphics[width=0.49\textwidth]{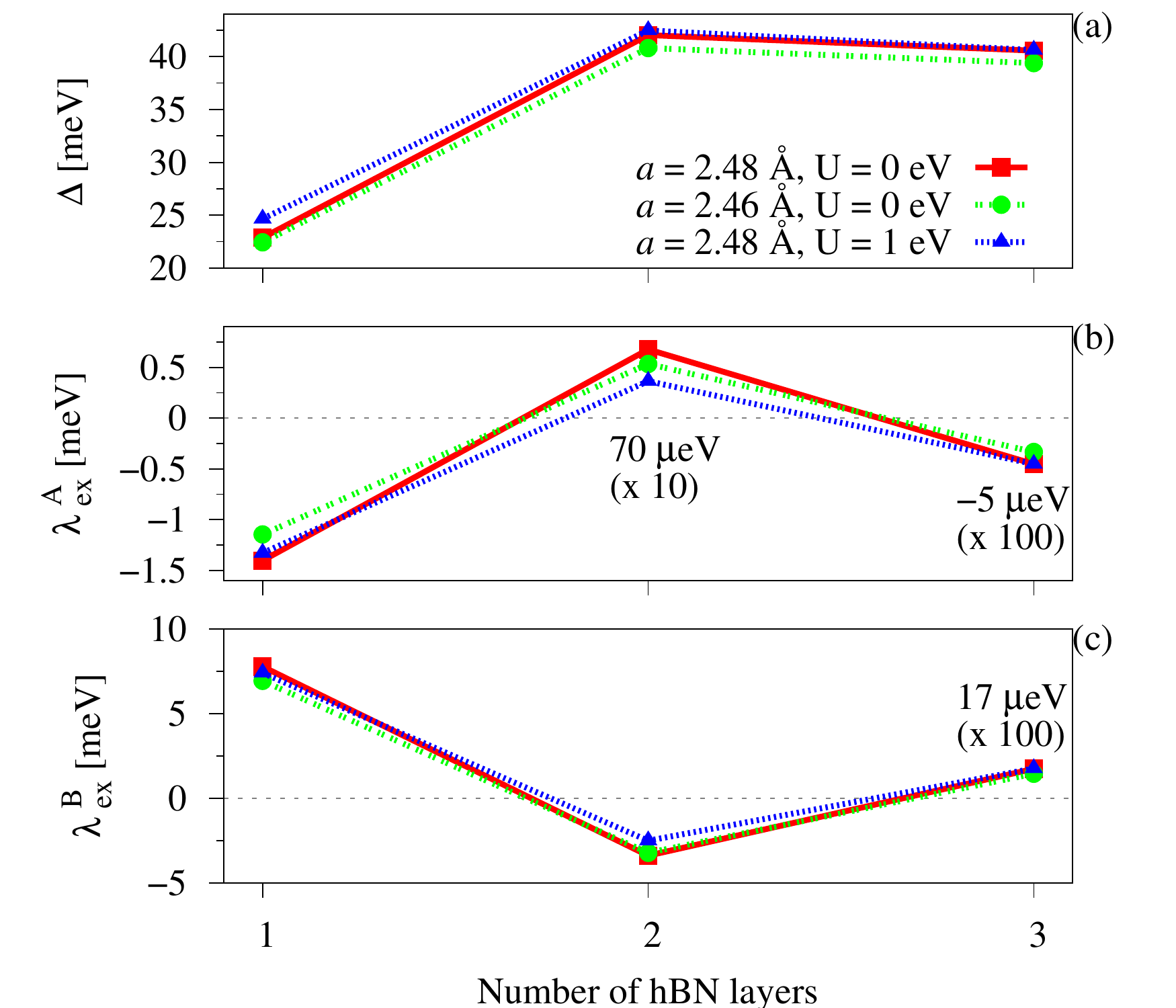}
 \caption{\label{fig:hBN_layers_Ni}(Color online) Influence of the number of hBN layers on the proximity-induced parameters for the graphene/hBN/Ni structure, using the $p_z$ model at the K point. Dependence of (a) the proximity gap $\Delta$, (b) the exchange parameters $\lambda_{\textrm{ex}}^{\textrm{A}}$, and (c) 
$\lambda_{\textrm{ex}}^{\textrm{B}}$ on the number of hBN layers for different lattice constants or an additional Hubbard parameter of $U = 1.0$~eV. Parameter values for two (three) layers of hBN were increased by a factor of 10 (100) for better visualization.}
 \end{figure}

Figure \ref{fig:hBN_layers_Ni} shows the dependence of the proximity gap $\Delta$ and the two exchange parameters $\lambda_{\textrm{ex}}^{\textrm{A}}$ and $\lambda_{\textrm{ex}}^{\textrm{B}}$ on the number of hBN layers between Ni and graphene. 
Again, similar to those for Co, the exchange parameters decrease by one order of magnitude and change sign after an additional insulating layer is added. 
The proximity gap $\Delta$ doubles for two layers of hBN and again stays constant since, effectively, the local environment for graphene does not change anymore the addition of hBN layers. 
(The distances for the three-layer case are similar to those for the two-layer case.) 
We have only one additional distance between the two hBN layers directly below graphene, which was relaxed to $d_{\textrm{hBN/hBN}} = 3.073$~${\textrm{\AA}}$.

\begin{figure}[htb] 
 \centering 
\includegraphics[width=0.49\textwidth]{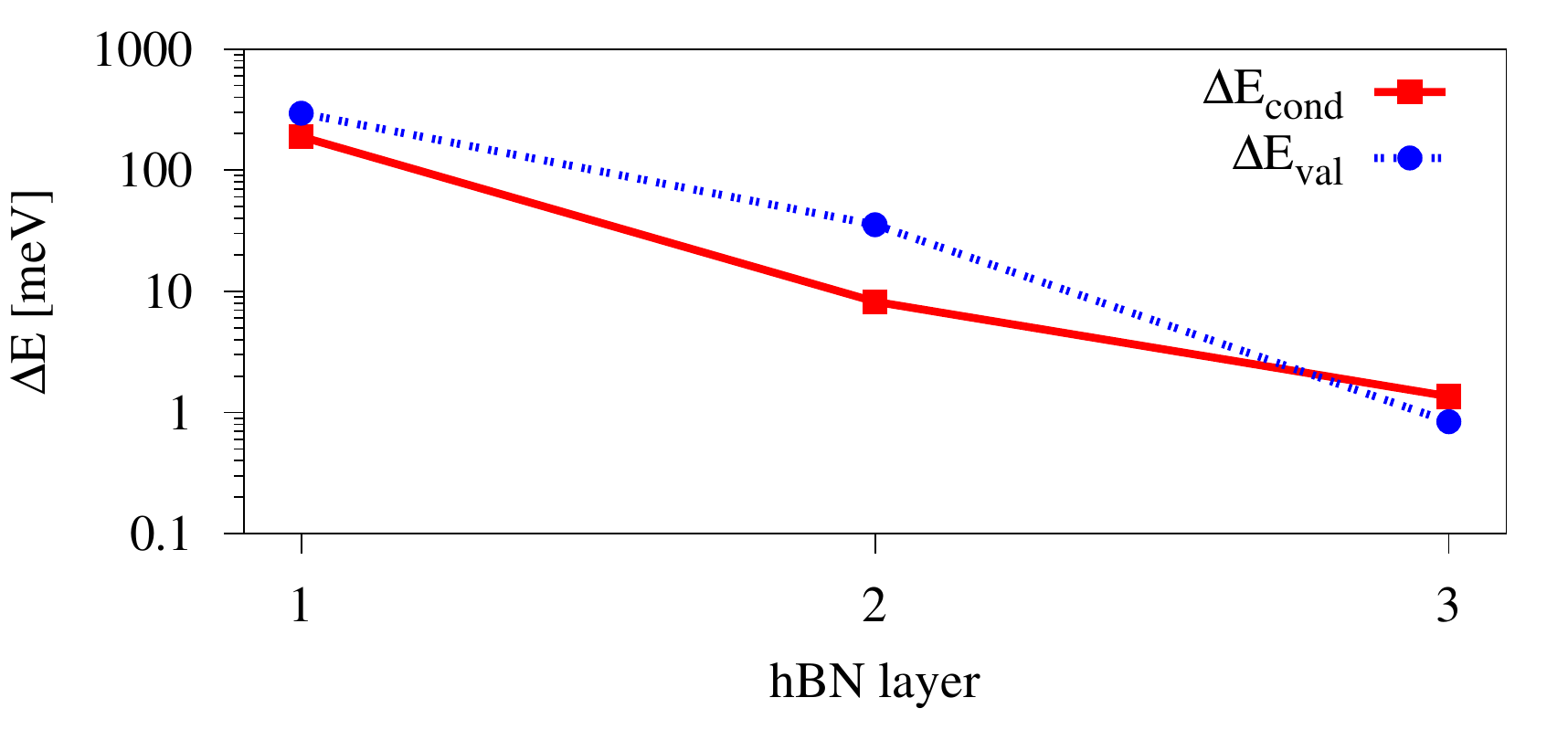}
 \caption{\label{fig:splittings_hBN_nickel}(Color online) Conduction $\Delta E_{\textrm{cond}}$ and valence $\Delta E_{\textrm{val}}$ band splittings of the three individual hBN layers at the K point. Values are obtained by identifying the spin-split hBN conduction and valence bands of the three individual layers in the band structure of graphene/hBN/Ni heterostructure for three layers of hBN.}
 \end{figure}

Also the bands of hBN are spin split, and like we did for the Co substrate, we look at the graphene/hBN/Ni structure with three layers of hBN. 
In the band structure we can identify the highest- (lowest-) lying valence (conduction) bands, which are spin split, of the three individual layers. From that, we extract the band splittings of conduction $\Delta E_{\textrm{cond}}$ and valence $\Delta E_{\textrm{val}}$ bands of the individual hBN layers at the K point. 
We notice that the spin-up bands of hBN are always lying lower in energy than the spin-down ones. 
In Fig. \ref{fig:splittings_hBN_nickel} we show the valence- and conduction-band splittings at the K point of the three layers. 
The splittings are very similar to, but smaller in magnitude than, those in the Co case. The spin splitting of the bands from the first layer is roughly $250$~meV.

\paragraph{Lattice-constant effects.}
We also look at how the band structure of the slabs changes when we use the graphene lattice constant for all the materials, $a = 2.46$~${\textrm{\AA}}$, by simply changing the in-plane lattice constant to this value without changing the vertical distances between the layers. 
The results in this case do not deviate much from the case with $a = 2.48$~${\textrm{\AA}}$, as can be seen in Fig. \ref{fig:hBN_layers_Ni}. 
The Fermi velocity for $a = 2.46$~${\textrm{\AA}}$~and one hBN layer is $v_{\textrm{F}} = 0.822\times 10^6$ m/s, corresponding to a larger nearest-neighbor hopping parameter of $t=2.52$~eV.
 
\paragraph{Hubbard U.}
We now introduce a Hubbard parameter $U = 1.0$~eV to compare 
the results of the calculations of different numbers of layers of hBN with 
the ones with $U=0$~eV. 
This comparison is in Fig. \ref{fig:hBN_layers_Ni}. 
In contrast to the case of Co, the proximity effects are barely affected by the positioning of the $d$ levels 
since the levels are quite far from the Dirac point. 
We can conclude that the predicted large proximity 
exchange splitting in the Dirac valence band is robust. 

\paragraph{Electric field effects.}

\begin{figure}[htb] 
 \centering 
 \includegraphics[width=0.49\textwidth]{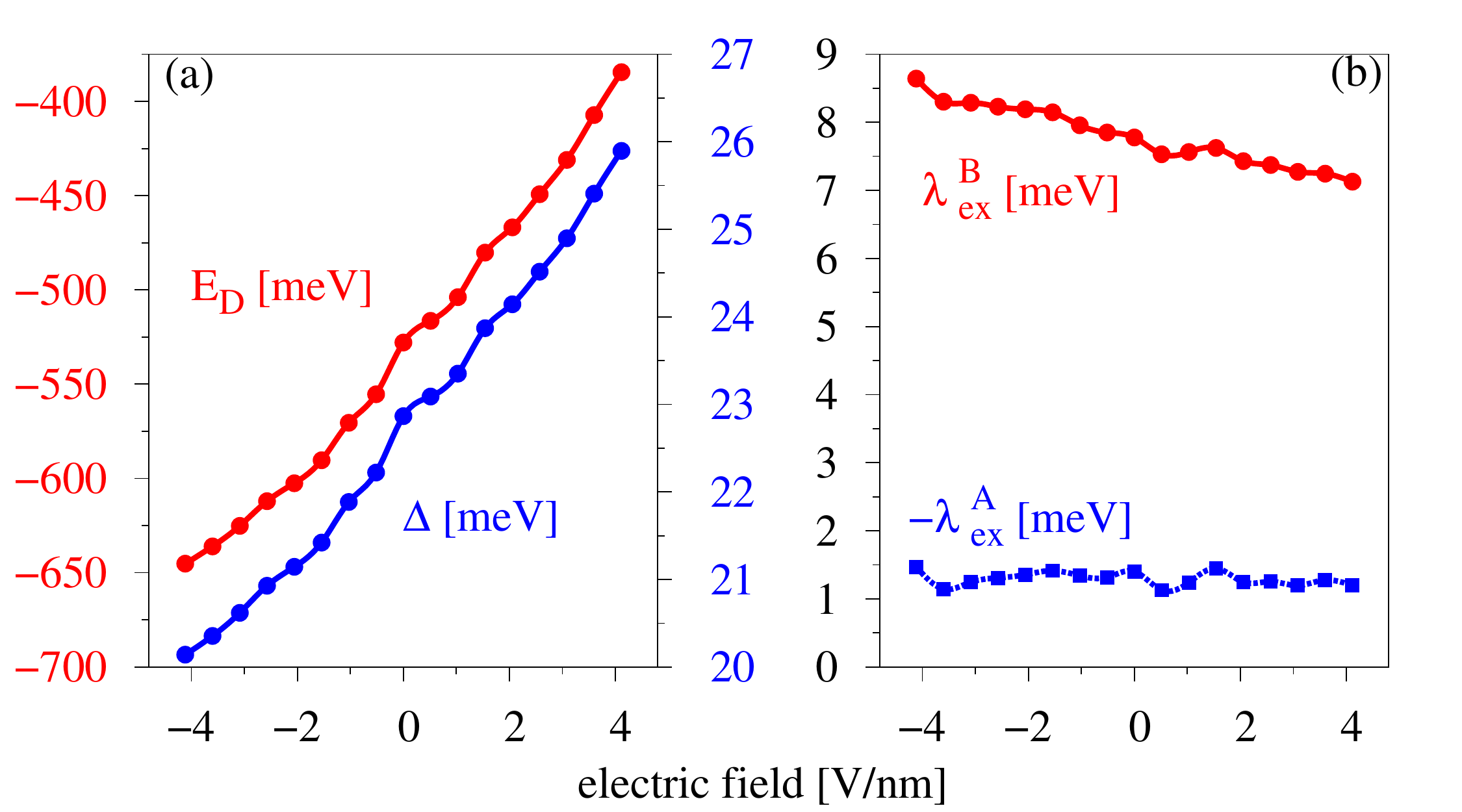}
 \caption{\label{fig:Efield_Ni}(Color online) Influence of the electric field on the proximity-induced parameters for the graphene/hBN/Ni structure for one hBN layer, using the $p_z$ model at the K point. Dependence of the (a) Dirac energy $E_{\textrm{D}}$ and the proximity gap $\Delta$ and (b) the exchange parameters $\lambda_{\textrm{ex}}^{\textrm{A}}$ and $\lambda_{\textrm{ex}}^{\textrm{B}}$ on the applied transverse electric field.}  
 \end{figure}

Figure \ref{fig:Efield_Ni} shows the influence of the electric field on the proximity parameters and the doping level. 
We can see that $E_{\textrm{D}}$ and $\Delta$ show the same trend with electric field. 
By increasing the electric field the doping level decreases. 
The proximity gap $\Delta$ also increases with increasing electric field, reflecting the charge transfer away from graphene. The continuous shift of the doping level with the applied electric field allows us to shift the Fermi level to the desired position.
Compared to those in the case of Co, the proximity parameters for Ni change more smoothly with applied electric field.
The magnitude of the proximity parameter $\lambda_{\textrm{ex}}^{\textrm{A}}$, on average, stays constant with electric field. 
The magnitude of the parameter $\lambda_{\textrm{ex}}^{\textrm{B}}$ slowly decreases
with electric field, but for moderate fields the band splittings are almost unchanged. The electric tunability of
the proximity exchange in this case is rather weak. 

 \begin{figure}[htb] 
 \centering 
 \includegraphics[width=0.49\textwidth]{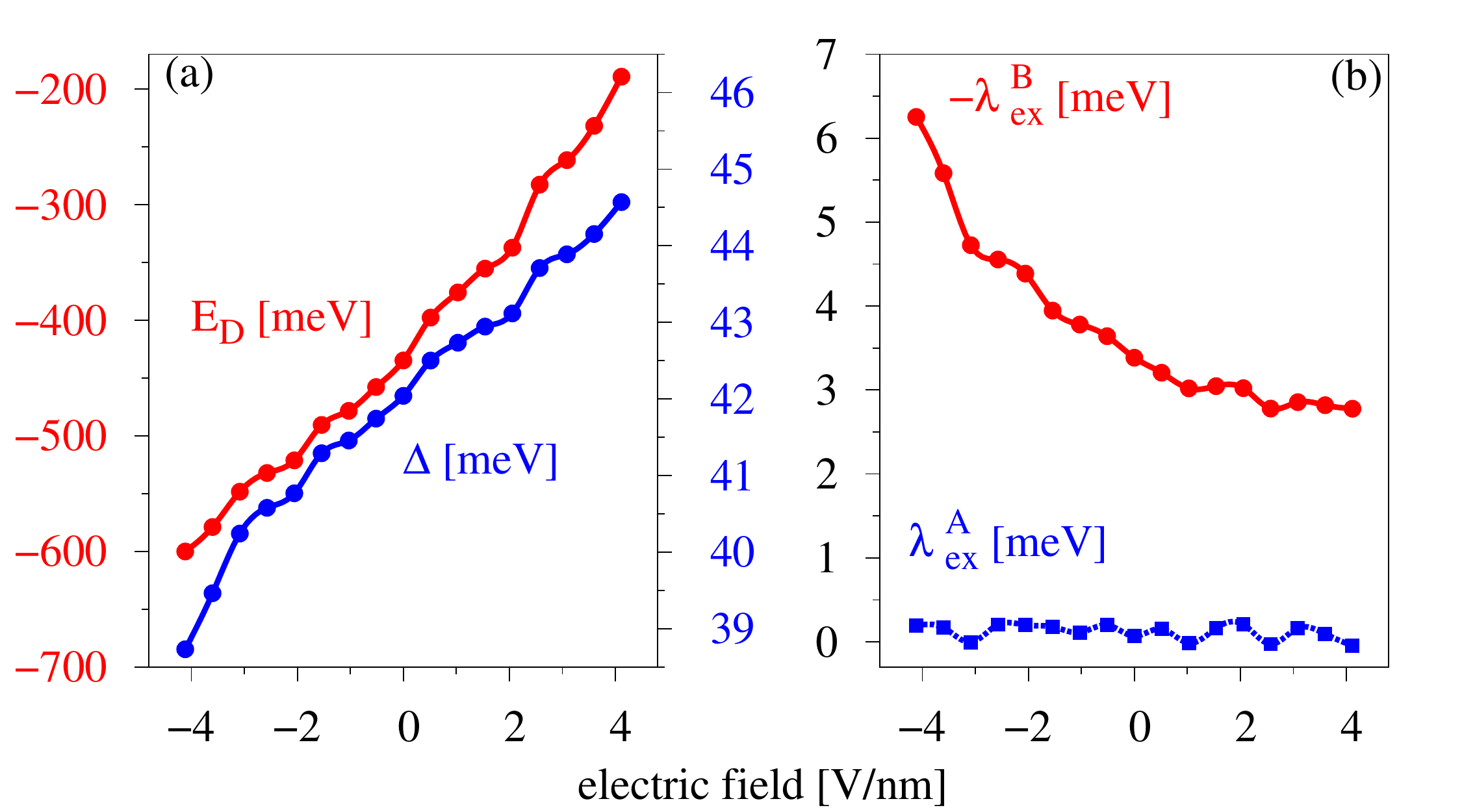}
 \caption{\label{fig:Efield_Ni_2hBN}(Color online) Influence of the electric field on the proximity-induced parameters for the graphene/hBN/Ni structure for two hBN layers, using the $p_z$ model at the K point. Dependence of the (a) Dirac energy $E_{\textrm{D}}$ and the proximity gap $\Delta$ and (b) the exchange parameters $\lambda_{\textrm{ex}}^{\textrm{A}}$ and $\lambda_{\textrm{ex}}^{\textrm{B}}$ on the applied transverse electric field. }  
 \end{figure} 

Figure \ref{fig:Efield_Ni_2hBN} shows the influence of the electric field on the proximity parameters and the doping level for two hBN layers.
We can see that $E_{\textrm{D}}$ and $\Delta$ show the same trend with electric field as for the single-layer hBN, but the orbital gap parameter $\Delta$ is roughly two times larger than in the case with monolayer hBN. 
As we have already seen, the two proximity parameters $\lambda_{\textrm{ex}}$ change their sign with the addition of the second hBN layer. 
The magnitude of the proximity parameter $\lambda_{\textrm{ex}}^{\textrm{A}}$ stays roughly constant with electric field but is one order of magnitude smaller than in the monolayer hBN case. 

The proximity parameter $\lambda_{\textrm{ex}}^{\textrm{B}}$ decreases with increasing electric field. For negative (positive) fields, the Dirac point is shifted in energy towards (away from) the hybridizing $d$ levels, which cross the valence Dirac states (see Fig. \ref{fig:Spin_bands_Ni_2hBN_Gr}), and $\lambda_{\textrm{ex}}^{\textrm{B}}$ is increasing (decreasing). We note that the magnitude of $\lambda_{\textrm{ex}}^{\textrm{B}}$ in the bilayer hBN case is comparable to that in the monolayer hBN case. 

\begin{figure}[htb] 
 \centering 
 \includegraphics[width=0.49\textwidth]{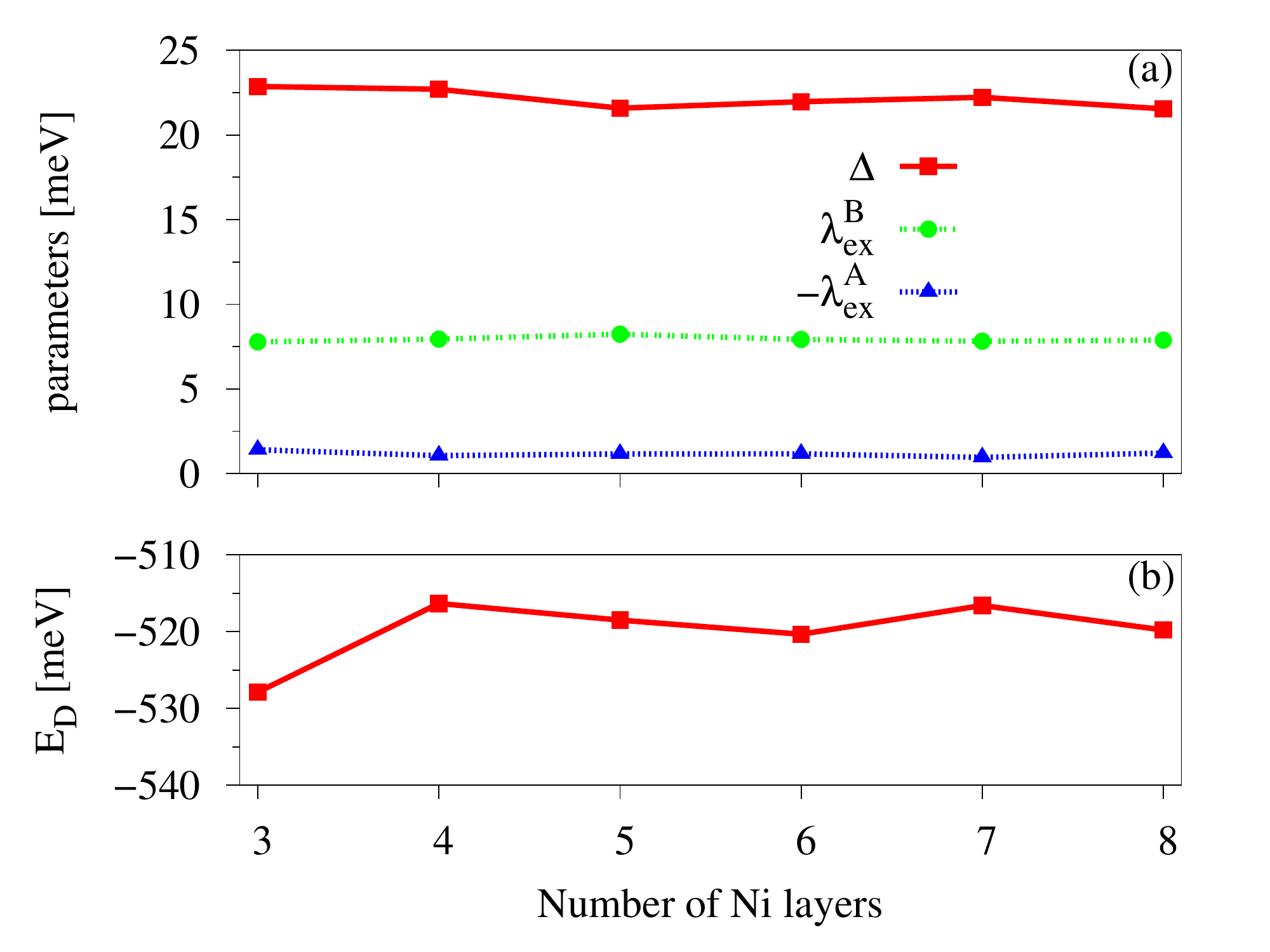}
\caption{\label{fig:Ni_layers}(Color online) Influence of the number of Ni layers on the band structure for the graphene/hBN/Ni system for one hBN layer, using the $p_z$-model at the K point. 
Dependence of the (a) the proximity gap $\Delta$ and the exchange parameters $\lambda_{\textrm{ex}}^{\textrm{A}}$ and $\lambda_{\textrm{ex}}^{\textrm{B}}$ and (b) the Dirac energy $E_{\textrm{D}}$ on the number of Ni layers.}  
 \end{figure}

\paragraph{Additional nickel layers.}
Finally, we analyze the influence of additional Ni layers on the band structure (see Fig. \ref{fig:Ni_layers}). As we increase the number of Ni layers, more $d$ bands are also introduced into the dispersion.
 Consequently, in the vicinity of the K point graphene Dirac states can be disturbed by these additional Ni bands. 
 We can see that the band splittings of graphene at the K point do not get influenced much by additional layers since the parameters stay at the same order.
In this case, already, four layers of Ni show a \textit{steady} situation for the Dirac energy $E_{\textrm{D}}$. The effect on the proximity parameters is negligible. 

\section{Summary and Conclusions}
\label{sec:Summary}

We investigated the proximity-induced exchange interaction induced by the ferromagnets Co and Ni into graphene through the insulator hBN. We found proximity-induced exchange splittings of up to $20$~meV together with a proximity gap of $40$~meV for one layer of hBN. 
As more insulating layers are introduced, the proximity-induced exchange interaction in general decreases exponentially, but the signs of the exchange parameters reverse. This reversal of the signs continues for 
up to four layers of hBN. 
We also introduced a minimal model and an extended model to fit the first-principles data. The model parameters 
are summarized in Tables \ref{tab:summary} and \ref{tab:summary_pz}.

A fascinating case is that of Co. Here a rather flat $d$ level strongly hybridizes with 
$p_z$ graphene orbitals in the valence band, leading to a giant proximity exchange in the case of two 
hBN layers. Since this giant exchange depends on the offset of the $d$ orbital energy and the Dirac point, 
we found that an external transverse electric field can tune this effect, and even lead to a crossover between
positive and negative induced spin polarization in the valence band of graphene. 
We found that in general the results for both ferromagnets are similar, although the effects of Co are stronger
than those of Ni, which is a consequence of the smaller atomic magnetic moment of the latter. 
The main difference between Co and Ni lies in the orbital decomposition of the $d$ bands, 
which interact with the graphene Dirac states, and leads to the giant spin splitting of the valence band
in the case of Co.

\section{Acknowledgments}
\label{sec:Acknowledgments}
This work was supported by DFG SFB Grant No. 689 and GRK Grant No. 1570 and by the EU Seventh Framework Programme under Grant Agreement No. 604391 Graphene Flagship.

\bibliography{LIB}

\end{document}